\documentclass[11pt]{article}
\usepackage{graphicx}
\usepackage{dcolumn}
\usepackage{array, longtable}
\usepackage{epsfig}
\usepackage{amsmath}
\usepackage{colordvi}
\usepackage{hhline}

\newcommand{\BaBarYear}{2006}

\newcommand{\BABARConfNumber} {037}
\newcommand{\SLACPubNumber}{12035}
\newcommand{\LANLNumber} {0608002}

\input pubboard/babarsym
\newcommand{\bei}{\begin{itemize}}
\newcommand{\eei}{\end{itemize}}
\newcommand{\beq}{\begin{equation}}
\newcommand{\eeq}{\end{equation}}
\newcommand{\beqn}{\begin{eqnarray}}
\newcommand{\eeqn}{\end{eqnarray}}
\newcommand{\beqns}{\begin{eqnarray*}}
\newcommand{\eeqns}{\end{eqnarray*}}

\newcommand{\intl}{\int\limits}

\newcommand{\e}{\epsilon}

\def\bk{\!\!\!\!}

\def\PRL{{\em Phys. Rev. Lett.}}

\def\ea{{\em et al.}}
\def\min{{\rm min}}

\def\rPTbarkappa{\kern 0.18em\overline{\kern -0.18em r}{}^{\kappa}{}}
\def\rPTbarsigma{\kern 0.18em\overline{\kern -0.18em r}{}^{\sigma}{}}

\def\deltabarkappa{\kern 0.18em\overline{\kern -0.18em \delta}{}_r^{\kappa}}
\def\deltabarsigma{\kern 0.18em\overline{\kern -0.18em \delta}{}_r^{\sigma}}
\def\deltaTbarkappa{\kern 0.18em\overline{\kern -0.18em \delta}{}_T^{\kappa}}
\def\deltaTbarsigma{\kern 0.18em\overline{\kern -0.18em \delta}{}_T^{\sigma}}

\newcommand\ph{\phantom}
\newcommand{\half}{\ensuremath{{1\over2}}}
\newcommand{\pvec}{{\bf p}}

\def\OC{X}
\def\OCbar{{\kern 0.18em\overline{\kern -0.18em \OC}}}
\def\mBz{m_{\Bz}}
\def\spz{s_{+}}
\def\smz{s_{-}}
\def\spm{s_{0}}

\def\mpm{m_{0}}

\def\fpz{f_{+}}
\def\fmz{f_{-}}
\def\fpm{f_{0}}

\def\mpmMax{\mpm^{\rm max}}
\def\mpmMin{\mpm^{\rm min}}

\def\pipipi{\pip\pim\piz}
\def\Btopipipi{\Bz\to\pipipi}

\def\mprime{m^\prime}
\def\thetaprime{\theta^\prime}
\def\deprime{{\de^\prime}{}}
\def\dt{\deltat}

\def\de{\DeltaE}
\def\demin{\de_{-}}
\def\demax{\de_{+}}
\def\deminmax{\de_{\pm}}

\def\tpi{3\pi}

\def\Qtag{q_{\rm tag}}
\def\Qtagi{q_{{\rm tag},i}}
\def\Atagqq{A_{q\bar q,\,\rm tag}}
\def\Atag{A_{{B^+,\,\rm tag}}}
\def\Atagj{A_{{B^+,\,\rm tag},j}}

\def\cat{c}

\def\detJ{|\det J|}
\def\detJi{|\det J_i|}
\def\Rscf{R_{\rm SCF}}

\def\a{\kappa}

\def\c{\tau}

\def\Amptp{{A}_{3\pi}}
\def\Amptpbar{\kern 0.18em\overline{\kern -0.18em {\cal A}}_{3\pi}}

\def\absAmptp{|\Amptp|}
\def\absAmptpbar{|\Amptpbar|}
\def\Amptpkappa{{A^{\kappa}}}
\def\Amptpsigma{{A^{\sigma}}}
\def\Amptpbarkappa{\kern 0.18em\overline{\kern -0.18em A}{}^{\kappa}{}}
\def\Amptpbarsigma{\kern 0.18em\overline{\kern -0.18em A}{}^{\sigma}{}}
\def\AmpAll{|{\cal A}_{3\pi}^\pm(\dmt)|^2}
\def\AmpAllp{|{\cal A}_{3\pi}^+(\dmt)|^2}
\def\AmpAllm{|{\cal A}_{3\pi}^-(\dmt)|^2}

\def\Tbarkappa{\kern 0.18em\overline{\kern -0.18em T}{}^{\kappa}{}}
\def\Tbarsigma{\kern 0.18em\overline{\kern -0.18em T}{}^{\sigma}{}}
\def\Pbarkappa{\kern 0.18em\overline{\kern -0.18em P}{}^{\kappa}{}}
\def\Pbarsigma{\kern 0.18em\overline{\kern -0.18em P}{}^{\sigma}{}}
\def\kappab{\overline\kappa}

\def\ji{\kappab}
\def\ij{\kappa}
\def\Aij{{A^{\ij}}}
\def\Abij{\kern 0.18em\overline{\kern -0.18em A}{}^{\ij}{}}
\def\Tij{{T^{\ij}}}
\def\Tji{{T^{\ji}}}
\def\Pij{{P^{\ij}}}
\def\Pji{{P^{\ji}}}
\def\R{{\rm Re}}
\def\I{{\rm Im}}
\def\kappm{\kappa^{+-}}
\def\kapmp{\kappa^{-+}}

\def\bbar{\Bz \Bzb}

\def\Crhopi{C}
\def\dCrhopi{\Delta C}

\def\dS{\Delta S}
\def\dC{\Delta C}
\def\rhopi{\rho\pi}
\def\Acp{{\cal A}_{\rho\pi}}

\def\Acppm{{\cal A}_{\rho\pi}^{+-}}
\def\Acpmp{{\cal A}_{\rho\pi}^{-+}}

\def\Nbpm{{\kern 0.18em\overline{\kern -0.18em N}}^{+-}}
\def\Nbmp{{\kern 0.18em\overline{\kern -0.18em N}}^{-+}}
\def\rar{\rightarrow}

\def\Mu{\mu}
\def\Chi2MinaMu{\chi^2_{\min ;\a,\Mu}}
\def\Chi2MinMu{\chi^2_{\min ;\Mu}(a)}

\def\dmt{\Delta t}
\def\dmd{\Delta m_d}

\def\TM{{\rm TM}}
\def\SCF{{\rm SCF}}

\def\fscfave{\kern 0.18em\overline{\kern -0.18em f}_{\rm SCF}}
\def\fscf{f_{\rm SCF}}
\def\fscfi{f_{{\rm SCF},i}}

\def\abar{\bar{a}}

\def\PRL{{\em Phys. Rev. Lett.}}

\def\Bbar{\kern 0.18em\overline{\kern -0.18em B}{}\xspace}

\def\BRpmb{{\cal \kern 0.18em\overline{\kern -0.18em  B}}{}_{\rho\pi}^{+-}}
\def\BRmpb{{\cal \kern 0.18em\overline{\kern -0.18em  B}}{}_{\rho\pi}^{-+}}

\def\BRipmb{{\cal \kern 0.18em\overline{\kern -0.18em  B}}{}_{\rho^+\pi^-}}
\def\BRimpb{{\cal \kern 0.18em\overline{\kern -0.18em  B}}{}_{\rho^-\pi^+}}

\def\Abar{\kern 0.18em\overline{\kern -0.18em A}{}}
\def\abar{\kern 0.18em\overline{\kern -0.18em a}{}}

\def\Apm{A^{+}}
\def\Amp{A^{-}}
\def\Apmb{\Abar^{+}}
\def\Ampb{\Abar^{-}}

\def\Azz{A^{0}}
\def\Azzb{\Abar^{0}}

\def\ie{{\em i.e.}} 
\def\cf{{\em cf.}} 
\def\eg{{\em e.g.}}

\newcommand{\UPm    }{\ensuremath{U_+^-}}
\newcommand{\UMp    }{\ensuremath{U_-^+}}
\newcommand{\UMm    }{\ensuremath{U_-^-}}
\newcommand{\UPMpRe }{\ensuremath{U_{+-}^{+,\R}}}
\newcommand{\UPMmRe }{\ensuremath{U_{+-}^{-,\R}}}
\newcommand{\UPMpIm }{\ensuremath{U_{+-}^{+,\I}}}
\newcommand{\UPMmIm }{\ensuremath{U_{+-}^{-,\I}}}
\newcommand{\Uzp    }{\ensuremath{U_0^+}}
\newcommand{\UPzpRe }{\ensuremath{U_{+0}^{+,\R}}}
\newcommand{\UPzpIm }{\ensuremath{U_{+0}^{+,\I}}}
\newcommand{\UMzpRe }{\ensuremath{U_{-0}^{+,\R}}}
\newcommand{\UMzpIm }{\ensuremath{U_{-0}^{+,\I}}}
\newcommand{\IP     }{\ensuremath{I_+}}
\newcommand{\IM     }{\ensuremath{I_-}}
\newcommand{\IPMRe  }{\ensuremath{I_{+-}^{\R}}}
\newcommand{\IPMIm  }{\ensuremath{I_{+-}^{\I}}}

\newcommand{\Iz     }{\ensuremath{I_{0}}}
\newcommand{\IMzIm  }{\ensuremath{I_{-0}^{\I}}}
\newcommand{\IMzRe  }{\ensuremath{I_{-0}^{\R}}}
\newcommand{\IPzIm  }{\ensuremath{I_{+0}^{\I}}}
\newcommand{\IPzRe  }{\ensuremath{I_{+0}^{\R}}}
\newcommand{\Uzm    }{\ensuremath{U_{0}^{-}}}
\newcommand{\UMzmRe }{\ensuremath{U_{-0}^{-,\R}}}
\newcommand{\UMzmIm }{\ensuremath{U_{-0}^{-,\I}}}
\newcommand{\UPzmRe }{\ensuremath{U_{+0}^{-,\R}}}
\newcommand{\UPzmIm }{\ensuremath{U_{+0}^{-,\I}}}

\newcommand{\sigPiNb}{\ensuremath{N_{\rm 3 \pi}}}

\long\def\inst#1{\par\nobreak\kern 4pt\nobreak
    {\it #1}\par\vskip 10pt plus 3pt minus 3pt}

\begin{document}

\begin{flushright}
\babar-CONF-\BaBarYear/\BABARConfNumber \\
SLAC-PUB-\SLACPubNumber \\
hep-ex/\LANLNumber \\
July 2006 \\
\end{flushright}
\par\vskip 5cm

\begin{center}
{\large \bf
\boldmath
Measurement of {\em CP}-Violating Asymmetries in
\boldmath$B^0\to(\rho\pi)^0$ \\
Using a Time-Dependent Dalitz Plot Analysis
} 
\end{center}

\bigskip

\begin{center}
\large The \babar\ Collaboration\\
\mbox{ }\\
\today
\end{center}
\bigskip \bigskip

\date{\today}

\begin{center}
\large \bf Abstract
\end{center}

We report a measurement of \CP-violating asymmetries in
$B^0\to(\rho\pi)^0\to\pi^+\pi^-\pi^0$ decays using a time-dependent
Dalitz plot analysis. The results are obtained from a data sample
of 347 million $\FourS \to B\Bbar$ decays, collected by the \babar\
detector at the \pep2\ asymmetric-energy \B~Factory at SLAC. 
We measure 26 coefficients of the bilinear form factor terms occurring
in the time-dependent decay rate of the \Bz meson and
 derive the physically relevant quantities
from these coefficients. 
In particular we find a three standard deviation evidence of direct
\CP-violation in
the $B^0\to\rho^\pm\pi^\mp$ decays, with systematic uncertainties
included. We also achieve a constraint of the angle $\alpha$
of the Unitarity Triangle. All results presented are preliminary.

\vfill
\begin{center}

Submitted to the 33$^{\rm rd}$ International Conference on High-Energy Physics, ICHEP 06,\\
26 July---2 August 2006, Moscow, Russia.

\end{center}
\vspace{1.0cm}
\begin{center}
{\em Stanford Linear Accelerator Center, Stanford University, 
Stanford, CA 94309} \\ \vspace{0.1cm}\hrule\vspace{0.1cm}
Work supported in part by Department of Energy contract DE-AC03-76SF00515.
\end{center}

\newpage
\begin{center}
\small

The \babar\ Collaboration,
\bigskip

%
{B.~Aubert,}
{R.~Barate,}
{M.~Bona,}
{D.~Boutigny,}
{F.~Couderc,}
{Y.~Karyotakis,}
{J.~P.~Lees,}
{V.~Poireau,}
{V.~Tisserand,}
{A.~Zghiche}
\inst{Laboratoire de Physique des Particules, IN2P3/CNRS et Universit\'e de Savoie,
 F-74941 Annecy-Le-Vieux, France }
% ...

\end{center}\newpage

\newpage

\setcounter{footnote}{0}


\section{INTRODUCTION}
\label{sec:Introduction}

Measurements of the parameter $\stwob$~\cite{BaBarSin2beta,BelleSin2beta} 
have established \CP violation in the $\Bz$ meson system 
and provide strong support for the Kobayashi and Maskawa model of 
this phenomenon as arising from a single phase in the three-generation
CKM quark-mixing matrix~\cite{CKM}. 
We present, in this paper, results from a time-dependent 
analysis of the $\Bz\to\pip\pim\piz$ Dalitz plot (DP) that is dominated 
by the $\rho(770)$ intermediate resonances of all charges and their 
interference. The goal of the analysis
is the simultaneous extraction of the strong transition 
amplitudes and the weak interaction phase 
$\alpha\equiv \arg\left[-V_{td}^{}V_{tb}^{*}/V_{ud}^{}V_{ub}^{*}\right]$
of the Unitarity Triangle. In the Standard Model, a non-zero 
value for $\alpha$ is responsible for the occurrence 
of mixing-induced \CP violation in this decay.
The \babar\  and Belle experiments have obtained constraints
on $\alpha$ from the measurement of effective quantities $\stwoa_{\rm eff}$
in $B$ decays to $\pip\pim$~\cite{babarpipi,bellepipi} and 
from $B$ decays to $\rho^+\rho^-$~\cite{babarrhorho,bellerhorho}, using an 
isospin analysis~\cite{GLisospin}.

Unlike $\pip\pim$, $\rho^{\pm}\pi^{\mp}$ is not a \CP 
eigenstate, and four flavor-charge configurations
$(\Bz(\Bzb) \to \rho^{\pm}\pi^{\mp})$ must be considered.  
The corresponding isospin analysis~\cite{Lipkinetal} is unfruitful
with the present statistics since two pentagonal amplitude relations with 
12 unknowns have to be solved (compared to 6 unknowns for
the  $\pip\pim$ and $\rho^+\rho^-$ systems). However, it
has been pointed out by Snyder and Quinn~\cite{SnyderQuinn}, that 
one can obtain the necessary degrees of freedom to constrain
$\alpha$ without ambiguity by explicitly including in the analysis the 
variation of the strong phases of the interfering $\rho$ resonances
in the Dalitz plot.

\subsection{DECAY AMPLITUDES}
\label{sec:kinmeatics}

We consider the decay of a spin-zero $\Bz$ with four-momentum
$p_B$ into the three daughters $\pip$, $\pim$, $\piz$,
with $p_+$, $p_-$, and $p_0$ their corresponding four-momenta. Using
as independent (Mandelstam) variables the invariant squared masses
\beq
\label{eq:dalitzVariables}
       \spz \;=\; (p_+ + p_0)^2~, \hspace{1cm}
       \smz \;=\; (p_- + p_0)^2~, 
\eeq
the invariant squared mass of the positive and negative pion, 
$\spm \;=\; (p_+ + p_-)^2$, is obtained from energy and 
momentum conservation
\beq
\label{eq:magicSum}
	\spm \;=\; \mBz^2 + 2m_{\pi^+}^2 + m_{\pi^0}^2
		   - \spz - \smz~.
\eeq
The differential $\Bz$ decay width with respect to the 
variables defined in Eq.~(\ref{eq:dalitzVariables}) (\ie, the 
{\em Dalitz plot}) reads
\beq
\label{eq:partialWidth}
	d\Gamma(\Btopipipi) \;=\; 
	\frac{1}{(2\pi)^3}\frac{|\Amptp|^2}{8 \mBz^3}\,d\spz d\smz~,
\eeq
where $\Amptp$ is the Lorentz-invariant amplitude
of the three-body decay. 

We assume in the following that the amplitudes $\Amptp$ and its complex 
conjugate $\Amptpbar$, corresponding to the transitions $\Bz\to\pip\pim\piz$ 
and $\Bzb\to\pip\pim\piz$, respectively, are dominated by the three 
resonances $\rho^+$, $\rho^-$ and $\rho^0$. The $\rho$ resonances
are assumed to be the sum of the ground state $\rho(770)$ and the
radial excitations $\rho(1450)$ and $\rho(1700)$, with 
resonance parameters determined by a combined fit 
to $\tau^+\to\nutb\pip\piz$ and $\epem\to\pip\pim$ data~\cite{taueeref}.
Since the hadronic environment is different in \B decays, we 
cannot rely on this result and therefore determine the relative $\rho(1450)$
and $\rho(1700)$
amplitudes simultaneously with the \CP parameters from the fit. Variations of
the other parameters and possible contributions to the $\Bz\to\pip\pim\piz$ 
decay other than the $\rho$'s are studied as part of the systematic 
uncertainties (Section~\ref{sec:Systematics}).

Including the $\BzBzb$ mixing parameter $q/p$ into the $\Bzb$ decay 
amplitudes, we can write~\cite{SnyderQuinn,BaBarPhysBook}
\beqn
\label{eq:amp}
   \Amptp    		
	&=& \fpz \Apm + \fmz \Amp + \fpm\Azz ~, \\
\label{eq:ampBar}
   \Amptpbar 	
	&=& \fpz \Apmb + \fmz \Ampb + \fpm\Azzb ~,
\eeqn
where the $f_\kappa$ (with $\kappa=\{+,-,0\}$ denote the charge of the
$\rho$ from the decay of the $\Bz$ meson) 
are functions of the Dalitz variables 
$\spz$ and $\smz$ that incorporate the kinematic and dynamical properties 
of the $\Bz$ decay into a (vector) $\rho$ resonance and a 
(pseudoscalar) pion, and where the $\Aij$ are 
complex amplitudes
that
may comprise weak and strong transition phases and that are independent 
of the Dalitz variables. Note that the definitions~(\ref{eq:amp})
and (\ref{eq:ampBar}) imply the assumption that the relative phases 
between the $\rho(770)$ and its radial excitations are \CP-conserving.

Following Ref.~\cite{taueeref}, the $\rho$
resonances are parameterized in  $f_\kappa$ by a modified relativistic Breit-Wigner 
function introduced by Gounaris and Sakurai (GS)~\cite{rhoGS}. Due to angular 
momentum conservation, the spin-one $\rho$ resonance is polarized 
in a helicity-zero state. For a $\rho^{\a}$ resonance with charge $\a$,
the GS function is multiplied by the kinematic function 
$-4|{\bf p}_\a||{\bf p}_\c|\cos\theta_{\a}$, where ${\bf p}_\a$ 
is the momentum of either of the daughters of $\rho$-resonance
defined in the $\rho$-resonance rest frame, and where
${\bf p}_\c$ is the momentum of the 
particle not from $\rho$ decay defined in the same frame,
and $\cos\theta_{\a}$ the cosine of the helicity angle of
the $\rho^{\a}$. For the $\rho^+$ ($\rho^-$), $\theta_{+}$ 
($\theta_{-}$) is defined by the angle between the $\pi^0$ ($\pi^-$) 
in the $\rho^+$ ($\rho^-$) rest frame and the $\rho^+$ ($\rho^-$) 
flight direction in the $\Bz$ rest frame. For the $\rho^0$, 
$\theta_{0}$ is defined by the angle between the $\pi^+$ in 
the $\rho^0$ rest frame and the $\rho^0$ flight direction in 
the $\Bz$ rest frame. With these definitions, each pair of GS functions 
interferes destructively at equal masses-squared.

The occurrence of $\cos\theta_{\a}$ 
in the kinematic functions substantially enhances the interference 
between the different $\rho$ bands in the Dalitz plot, and thus 
increases the sensitivity of this analysis~\cite{SnyderQuinn}.

\subsection{TIME DEPENDENCE}

With $\deltat \equiv t_{\tpi} - t_{\rm tag}$ defined as the proper 
time interval between the decay of the fully reconstructed $B^0_{\tpi}$ 
and that of the  other meson $\Bz_{\rm tag}$,  the time-dependent decay
rate $\AmpAllp$ ($\AmpAllm$) when the tagging meson is a $\Bz$ ($\Bzb$) 
is given by 
\beq
\label{eq:dt}
    \AmpAll
	=
		\frac{e^{-|\dmt|/\tau_{B^0}}}{4\tau_{B^0}}
	\bigg[\absAmptp^2 + \absAmptpbar^2
	      \mp \left(\absAmptp^2 - \absAmptpbar^2\right)\cos(\dmd\dmt)
	      \pm\,2\I\left[\Amptpbar\Amptp^*\right]\sin(\dmd\dmt)	
	\bigg]~,
\eeq
where $\tau_{B^0}$ is the mean \Bz lifetime and $\deltamd$ is the $\BzBzb$ 
oscillation frequency. Here, we have assumed that \CP violation in $\bbar$ 
mixing is absent ($|q/p|=1$), $\Delta\Gamma_{B_d}=0$ and \CPT is conserved.
Inserting the amplitudes~(\ref{eq:amp}) and (\ref{eq:ampBar}), one 
obtains for the terms in Eq.~(\ref{eq:dt})
\beqn
   \label{eq:UI}
   \absAmptp^2 \pm \absAmptpbar^2 
	&=&
	\sum_{\kappa\in\{+,-,0\}}  |f_\kappa|^2U_\kappa^\pm
	\;\;+ \;\;
	2\bk\bk\sum_{\kappa <\sigma\in\{+,-,0\}} 
	\left(
	    \,\R\left[f_\kappa f_\sigma^*\right]U_{\kappa\sigma}^{\pm,\R}
	  - \,\I\left[f_\kappa f_\sigma^*\right]U_{\kappa\sigma}^{\pm,\I}
	\right)~,
	\nonumber\\[0.2cm]
   \I\left(\Amptpbar\Amptp^*\right)
	&=&
	\sum_{\kappa\in\{+,-,0\}}  |f_\kappa|^2I_\kappa
	\;\;+ 
	\sum_{\kappa <\sigma\in\{+,-,0\}} 
	\left(
	    \,\R\left[f_\kappa f_\sigma^*\right]I_{\kappa\sigma}^{\I}
	  + \,\I\left[f_\kappa f_\sigma^*\right]I_{\kappa\sigma}^{\R}
	\right)~,	
\eeqn
with
\beqn
\label{eq:firstObs}
   U_\kappa^\pm 		&=& |\Amptpkappa|^2 \pm |\Amptpbarkappa|^2~, \\
   U_{\kappa\sigma}^{\pm,\R(\I)}&=& \R(\I)\left[\Amptpkappa \Amptpsigma{}^* 
				    \pm \Amptpbarkappa \Amptpbarsigma{}^*\right]~, \\
   I_\kappa			&=& \I\left[\Amptpbarkappa\Amptpkappa{}^*\right]~, \\
   I_{\kappa\sigma}^{\R}  	&=& \R\left[\Amptpbarkappa\Amptpsigma{}^* 
					    - \Amptpbarsigma\Amptpkappa{}^*\right]~, \\
\label{eq:lastObs}
   I_{\kappa\sigma}^{\I}  	&=& \I\left[\Amptpbarkappa\Amptpsigma{}^* 
					    + \Amptpbarsigma\Amptpkappa{}^*\right]~.
\eeqn

The 27 coefficients~(\ref{eq:firstObs})--(\ref{eq:lastObs}) are real-valued
parameters that multiply the $f_\kappa f_\sigma^*$ bilinears (where $\kappa$
and $\sigma$ denote the charge of the $\rho$ resonances)~\cite{quinnsilva}. 
These are the 
observables that are determined by the fit. Each of the coefficients 
is related in a unique way to physically more intuitive quantities,
such as tree-level and penguin-type amplitudes, the angle $\alpha$, or
the quasi-two-body \CP and dilution parameters~\cite{rhopipaper} 
(\cf\   Section~\ref{sec:Physics}). The parameterization~(\ref{eq:UI}) is 
general; the information on the mirror solutions (\eg, on the angle $\alpha$)
that are present in the transition amplitudes $\Aij$, $\Abij$ is conserved.
In this paper, we determine the relative values of $U$ and $I$ coefficients 
to $U_+^+$.

The choice to fit for the $U$ and $I$ coefficients rather than
fitting for the complex transition amplitudes and the weak phase $\alpha$
directly is motivated by the following technical
simplifications: $(i)$ in contrast to the amplitudes, there is a unique
solution for the $U$ and $I$ coefficients requiring only a single fit
to the selected data sample,
$(ii)$ in the presence of background, the $U$ and $I$ coefficients are
approximately Gaussian distributed, which in general is not the case
for the amplitudes, and $(iii)$ the propagation of systematic uncertainties
and the averaging between different measurements are straightforward for
the $U$'s and $I$'s.

The $U_\kappa^+$ coefficients are related to resonance fractions (branching
fractions and charge asymmetries); the $U_\kappa^-$ determine the relative 
abundance of the $\Bz$ decay into $\rho^+\pim$ and $\rho^-\pip$ 
and the time-dependent direct \CP asymmetries. The $I_\kappa$ measure 
mixing-induced \CP violation and are sensitive to strong phase shifts.
Finally, the $U_{\kappa\sigma}^{\pm,\R(\I)}$ and $I_{\kappa\sigma}^{\R(\I)}$
describe the interference pattern in the Dalitz plot, and their presence
distinguishes this analysis from the quasi-two-body analysis previously
reported in~\cite{rhopipaper}.
They represent the additional degrees of freedom that allow one to 
determine the unknown penguin pollution and the relative strong phases.
However, because 
the overlap regions of the resonances are small and because the events
reconstructed in these regions suffer from large misreconstruction 
rates and background, a substantial data sample is needed to perform a 
fit that constrains all amplitude parameters.

We determine the quantities of interest in a subsequent least-squares
fit to the measured $U$ and $I$ coefficients.

\subsection{NORMALIZATION}

The decay rate~(\ref{eq:dt}) is used as a probability density 
function (PDF) in a maximum-likelihood fit and must therefore be normalized:
\beq
 	\AmpAll \;\longrightarrow\;
	\frac{1}{\langle |\Amptp|^2 + |\Amptpbar|^2 \rangle }\AmpAll~,
\eeq
where
\beqn
\label{eq:Norm}
	\langle |\Amptp|^2 + |\Amptpbar|^2 \rangle
	\;=\;
	\sum_{\kappa\in\{+,-,0\}}  \langle|f_\kappa|^2\rangle U_\kappa^+
	\;+\;
	2\R\!\!\!\!\!\!\!\!
		\sum_{\kappa <\sigma\in\{+,-,0\}} \!\!\!\!
			\langle f_\kappa f_\sigma^*\rangle
			\left(
				U^{+,\mathrm{Re}}_{\kappa\sigma} +
				i\cdot U^{+,\mathrm{Im}}_{\kappa\sigma}
			\right)
	   ~.
\eeqn
The complex expectation values $\langle f_\kappa f_\sigma^*\rangle$ are 
obtained from high-statistics Monte Carlo integration of the Dalitz 
plot~(\ref{eq:partialWidth}), taking into account acceptance and 
resolution effects.

The normalization of the decay rate~(\ref{eq:dt}) renders the normalization
of the $U$ and $I$ coefficients arbitrary, so that we can fix one 
coefficient. By convention, we set $U_+^+\equiv1$.

\subsection{THE SQUARE DALITZ PLOT}
\label{sec:SquareDP}

\begin{figure}[b]
  \centerline{ \epsfxsize8.2cm\epsffile{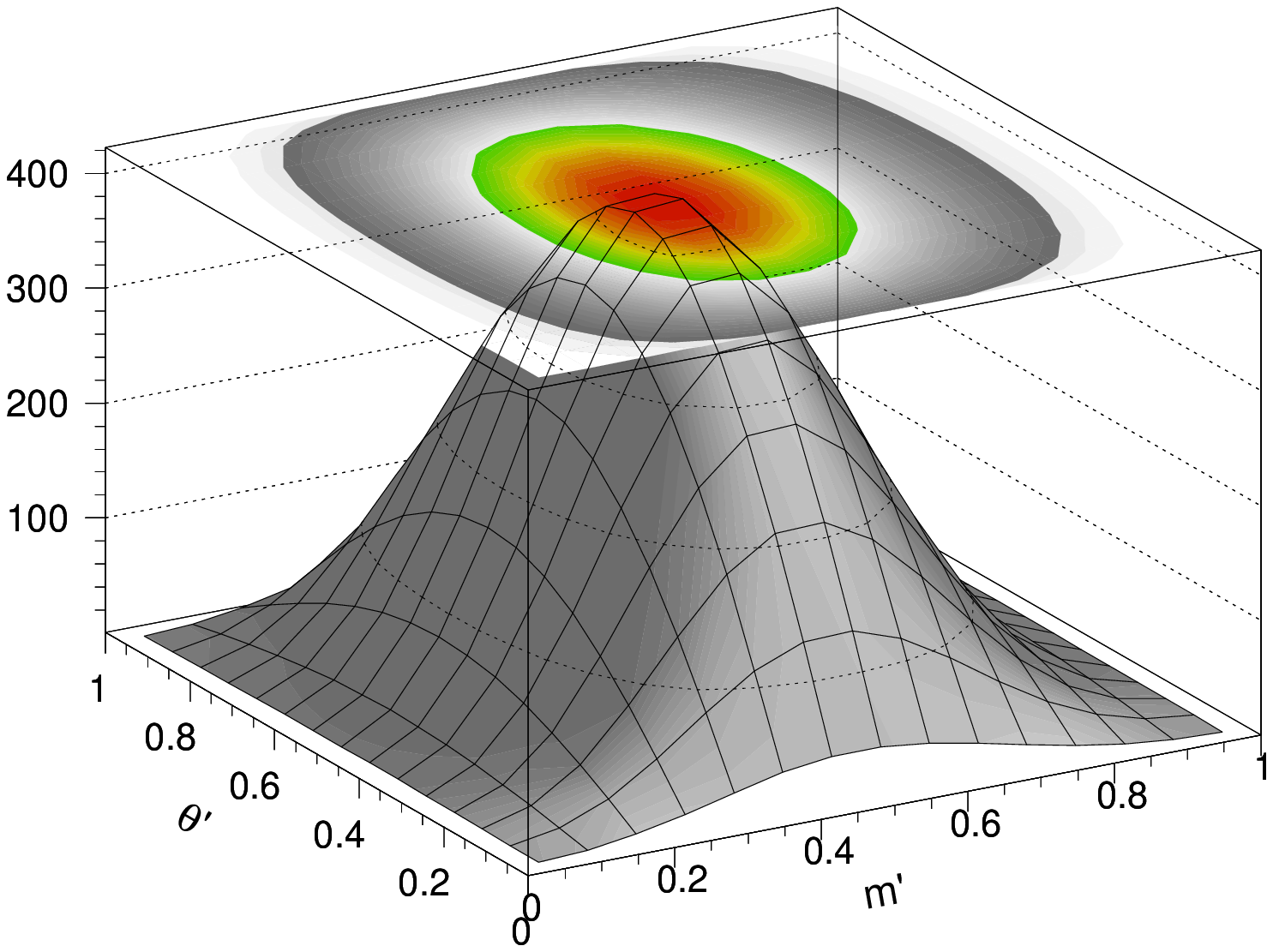}}
  \caption{\label{fig:jacobian}
	Jacobian determinant~(\ref{eq:detJ}) of the 
	transformation~(\ref{eq:SqDalitzTrans}) defining the SDP.
	Such pattern would be obtained in the SDP if events were uniformly
	distributed over the nominal Dalitz plot.}
\end{figure}

\begin{figure*}[t]
  \centerline{\epsfxsize8.2cm\epsffile{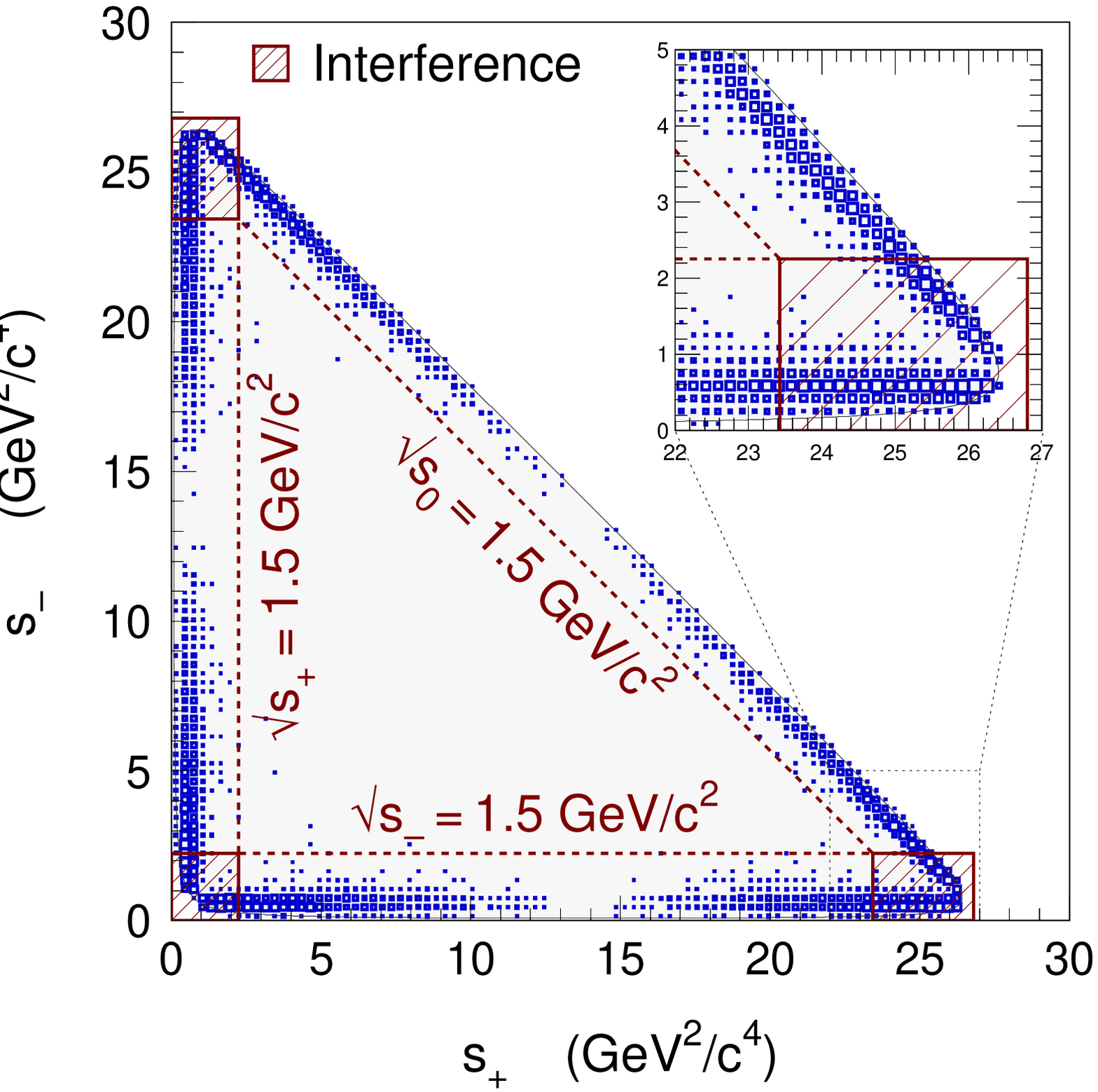}
	      \epsfxsize8.2cm\epsffile{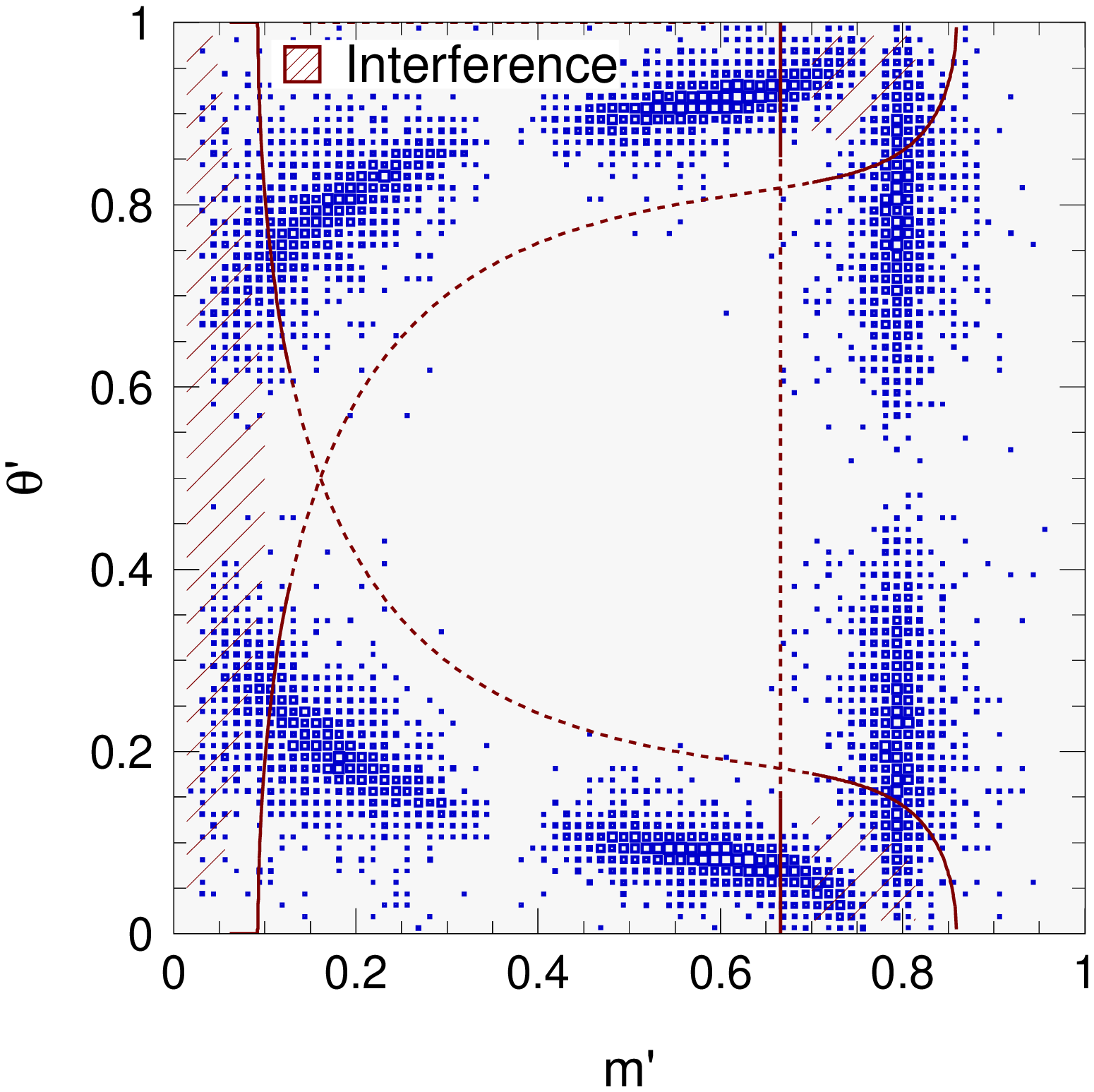}}
  \caption{\label{fig:DPs}
	Nominal (left) and square (right) Dalitz plots for Monte-Carlo
	generated $\Bz\rar\pi^+\pi^-\pi^0$ decays. Comparing the two
        Dalitz plots shows that the transformation~(\ref{eq:SqDalitzTrans})
        indeed homogenizes the distribution of events which are no more
  	along the plot boundaries but rather cover a larger fraction of
	the physical region. The decays have been
	simulated without any detector effect and the amplitudes $\Apm$,
	$\Amp$ and $\Azz$ have all been chosen equal to 1 in order to
	have destructive interferences at equal $\rho$ masses. The main
	overlap regions between the charged and neutral $\rho$ bands
	are indicated by the hatched areas.
	Dashed lines in both plots correspond to
	$\sqrt{s_{+,-,0}}=1.5~{\rm GeV}/c^2$: the central region of the Dalitz
	plot (defined by requiring that all 3 two-body invariant masses exceed
	this threshold) contains almost no signal event.}
\end{figure*}

Both the signal events and the combinatorial $\epem\to q\bar q$ ($q=u,d,s,c$) 
continuum background events populate the kinematic boundaries of the 
Dalitz plot due to the low final state masses compared to the $\Bz$ mass. 
We find the representation~(\ref{eq:partialWidth}) is inadequate when one wants 
to use empirical reference shapes in a maximum-likelihood fit.
Large variations occurring in small areas of the Dalitz plot are very difficult to describe in detail.  These regions are particularly important since this is where the interference, and hence our ability to determine the strong phases, occurs. 
We therefore apply the transformation
\beq
\label{eq:SqDalitzTrans}
	d\spz \,d\smz \;\longrightarrow \detJ\, d\mprime\, d\thetaprime~,
\eeq
which defines the {\em Square Dalitz plot} (SDP). The new coordinates 
are
\beq
\label{eq:SqDalitzVars}
	\mprime \equiv \frac{1}{\pi}
		\arccos\left(2\frac{\mpm - \mpmMin}{\mpmMax - \mpmMin}
			- 1
		      \right),~
	\thetaprime \equiv \frac{1}{\pi}\theta_{0},
\eeq
where $\mpm$ is the invariant mass between the charged tracks,
$\mpmMax=\mBz - m_{\pi^0}$ and $\mpmMin=2m_{\pi^+}$ are the kinematic
limits of $\mpm$, $\theta_{0}$ is the $\rho^0$ helicity angle,
and $J$ is the Jacobian of the transformation 
that zooms into the kinematic boundaries of the Dalitz plot.
The new variables range between 0 and 1.
The determinant of the Jacobian is given by
\beq
\label{eq:detJ}
	\detJ \;=\;	4 \,|{\bf p}^*_+||{\bf p}^*_0| \,\mpm
			\cdot 	
			\frac{\partial \mpm}{\partial \mprime}
			\cdot 	
			\frac{\partial \cos\theta_{0}}{\partial \thetaprime}~,
\eeq
where 
$|{\bf p}^*_+|=\sqrt{E^*_+ - m_{\pi^+}^2}$ and
$|{\bf p}^*_0|=\sqrt{E^*_0 - m_{\pi^0}^2}$, and where the energies 
$E^*_+$ and $E^*_0$ are in the $\pi^+\pi^-$ rest frame.
Figure~\ref{fig:jacobian} shows the determinant of the Jacobian as a function
of the SDP parameters $\mprime$ and $\thetaprime$. If the events in the
nominal Dalitz plot were distributed according to a uniform (non-resonant)
prior, their distribution in the SDP would match the plot of $\detJ$.

The effect of the transformation~(\ref{eq:SqDalitzTrans}) can be seen by
looking at Figure~\ref{fig:DPs} which displays the nominal and square Dalitz
plots for signal events generated with toy Monte Carlo (MC): the homogenization
of the distribution is clearly visible. This simulation does not take into
account any detector effect and corresponds to a particular choice of the
decay amplitudes for which destructive interferences occur at equal $\rho$
masses. To simplify the comparison, hatched areas showing the 
interference regions between $\rho$ bands and dashed isocontours
$\sqrt{s_{+,-,0}}=1.5~{\rm GeV}/c^2$ have been superimposed on both Dalitz
plots.

\section{THE \babar\ DETECTOR AND DATASET}
\label{sec:babar}

The data used in this analysis were collected with the \babar\ 
detector at the \pep2\ asymmetric-energy $e^+e^-$ storage ring at 
SLAC between October 1999 and June 2006. The sample consists of about
$310\;\mathrm{fb}^{-1}$, corresponding to $(346\pm3)\times10^{6}$ 
$B\Bbar$ pairs collected at the \FourS resonance (``on-resonance''), 
and an integrated luminosity of 21.6~\invfb collected about 40~\mev 
below the~\FourS (``off-resonance'').

A detailed description of the \babar\ detector is presented in 
Ref.~\cite{babarNim}. The tracking system used for track and vertex 
reconstruction has two main components: a silicon vertex tracker 
(SVT) and a drift chamber (DCH), both operating within a 1.5~T 
magnetic field generated by a superconducting solenoidal magnet. 
Photons are identified in an electromagnetic calorimeter (EMC) 
surrounding a detector of internally reflected Cherenkov light 
(DIRC), which associates Cherenkov photons with tracks for particle 
identification (PID). Muon candidates are identified with the
use of the instrumented flux return (IFR) of the solenoid.

\section{ANALYSIS METHOD}
\label{sec:Analysis}

The $U$ and $I$ coefficients and the $\Btopipipi$ event yield are
determined by a maximum-likelihood fit of the signal model to the 
selected candidate events. Kinematic and event shape variables 
exploiting the characteristic properties of the events are used 
in the fit to discriminate signal from background. 

\subsection{EVENT SELECTION AND BACKGROUND SUPPRESSION}
\label{subsec:selection}

We reconstruct $\Btopipipi$ candidates from pairs of 
oppositely-charged tracks, which are required to form a good quality vertex,
and a $\pi^0$ candidate.  In order to ensure that all events are within 
the Dalitz plot bounaries, we constrain the three-pion invariant mass to the B-mass.  
We use information from the tracking system, EMC, and DIRC to 
remove tracks for which the PID is consistent with the electron, kaon, 
or proton hypotheses. In addition, we require that at least one track 
has a signature in the IFR that is inconsistent with the muon 
hypothesis.
The $\pi^0$ candidate mass must satisfy $0.11<m(\gamma\gamma)<0.16\gevcc$, 
where each photon is required to have an energy greater than $50\mev$
in the laboratory frame (LAB) and to exhibit a lateral profile of energy 
deposition in the EMC consistent with an electromagnetic shower.

A $B$-meson candidate is characterized kinematically by the energy-substituted 
mass $\mes=\lbrack{(\half s+\pvec_0\cdot\pvec_B)^2/E_0^2-\pvec_B^2}\rbrack^\half$
and energy difference $\de = E_B^*-\half\sqrt{s}$, 
where $(E_B,\pvec_B)$ and $(E_0,\pvec_0)$ are the four-vectors
of the $B$-candidate and the initial electron-positron system,
respectively. The asterisk denotes the \FourS\  frame,
and $s$ is the square of the invariant mass of the electron-positron system.  
We require $5.272 < \mes <5.288\gevcc$, which retains $81\%$
of the signal and $8\%$ of the continuum background events. 
The $\de$ resolution 
exhibits a dependence on the $\pi^0$ energy and therefore varies 
across the Dalitz plot. We account for this effect by introducing
the transformed quantity $\deprime=(2\de - \demax - \demin)/(\demax - \demin)$,
with $\deminmax(\mpm)=c_{\pm}-\left(c_{\pm}\mp\bar c\right)(\mpm/\mpmMax)^2$,
where $\mpm$ is strongly correlated with the energy of $\piz$. 
We use the values
$\bar c = 0.045\gev$, $c_{-} = -0.140\gev$, $c_{+} = 0.080\gev$,
$\mpmMax = 5.0\gev$, and require $-1<\deprime<1$. 
These values have been obtained from Monte Carlo simulation and
are tuned to maximize the selection of correctly reconstructed over 
misreconstructed signal events. The requirement retains $75\%$ ($25\%$)
of the signal (continuum).

Backgrounds arise primarily from random combinations in continuum events.
To enhance discrimination between signal and continuum, we 
use a neural network (NN)~\cite{NNo} to combine four discriminating variables: 
the angles with respect to the beam axis of the $B$ momentum and $B$ thrust 
axis in the \FourS\ frame, and the zeroth and second order monomials
$L_{0,2}$ of the energy flow about the $B$ thrust axis.  The monomials
are defined by $ L_j = \sum_i {\bf p}_i\times\left|\cos\theta_i\right|^j$,
where $\theta_i$ is the angle with respect to the $B$ thrust axis of
track or neutral cluster $i$, ${\bf p}_i$ is its momentum, and the sum
excludes the $B$ candidate.  
The NN is trained in the signal region with off-resonance data and
simulated signal events. The final sample of signal candidates 
is selected with a requirement on the NN output that retains $77\%$ ($8\%$) 
of the signal (continuum).

The time difference $\deltat$ is obtained from the measured distance between 
the $z$ positions (along the beam direction) of the $\Bz_{\tpi}$ and 
$\Bz_{\rm tag}$ decay vertices, and the boost $\beta\gamma=0.56$ of 
the \epem\ system: $\deltat = \Delta z/\beta\gamma c$.
To determine the flavor of the $\Bz_{\rm tag}$ 
we use the $B$ flavor tagging algorithm of Ref.~\cite{BabarS2b}.
This produces six mutually exclusive tagging categories. We also 
retain untagged events in a seventh category to improve the efficiency 
of the signal selection and because these events contribute to the 
measurement of direct \CP violation. Events with multiple \B 
candidates passing the full selection occur 
in $16\%$ $(\rho^\pm\pi^\mp)$ and $9\%$ $(\rho^0\pi^0)$ 
of the cases. If the multiple candidates have different $\pi^0$'s, 
we choose the candidate with the reconstructed $\pi^0$ mass closest 
to the nominal one; in the case that both candidates have the same $\pi^0$,
we pick the first one.

The signal efficiency determined from MC simulation is $24\%$ for 
$B^0 \to \rho^\pm\pi^\mp$ and $B^0 \to \rho^0\pi^0$ events, and 
$11\%$ for non-resonant $\Btopipipi$ events. 

Of the selected signal events, $22\%$ of $B^0 \to \rho^\pm\pi^\mp$, 
$13\% $ of $B^0 \to \rho^0\pi^0$, and $6\%$ of non-resonant events are 
misreconstructed.  Misreconstruced events occur when a track or 
neutral cluster from the tagging $B$ is assigned to the reconstructed signal candidate. 
This occurs most often for  low-momentum tracks and photons and hence the misreconstructed events 
concentrate in the corners of the Dalitz plot.  Since these are also the areas where the $\rho$-mesons
overlap strongly, it is important to model the misreconstruced events correctly.  The details of the model
for misreconstructed events over the Dalitz plot is detailed in Section \ref{sec:deltaT}.

\subsection{BACKGROUND FROM OTHER {\em B} DECAYS}

\begin{table*}[t]
\begin{center}
\setlength{\tabcolsep}{0.0pc}
\begin{tabular*}{\textwidth}{@{\extracolsep{\fill}}llrr}
\hline
Class		& Mode								& BR~$[10^{-6}]$		& Expected number of events \\
\hline\\[-0.3cm]
0	& $B^+ \rar \rho^+\rho^0_{\rm\:[long]}$					& $   19.1 \pm\ph{}   3.5$ 	& $   52 \pm\ph{}    10$ \\
0	& $B^+ \rar a_1^+ (\rar (\rho\pi)^+) \piz$ 				& $   20.0 \pm\ph{}   15.0$ 	& $   32 \pm\ph{}   24$ \\
0	& $B^+ \rar a_1^0 (\rar \rho^{+-} \pi^{-+}) \pi^+$ 			& $   20.0 \pm        15.0$ 	& $   19 \pm\ph{}    14$ \\
1	& $B^+ \rar \pi^+\rho^0$ 						& $    8.7 \pm\ph{0}   1.0$ 	& $   73 \pm\ph{0}    8$ \\
1	& $B^+ \rar \rho^0 K^+$ 						& $    4.3 \pm\ph{0}   0.6$ 	& $    6 \pm\ph{0}    1$ \\
2	& $B^+ \rar \pi^+K^0_S(\rar\pi^+\pi^-)$ 				& $   8.3 \pm\ph{0}    0.4$ 	& $    10 \pm\ph{0}    1$ \\
3	& $B^+ \rar \pi^0\rho^+$ 						& $   10.8 \pm\ph{0}    1.4$ 	& $   63 \pm\ph{0}    8$ \\
3	& $B^+\rar \pi^+K^0_S(\rar \pi^0\pi^0)$ 				& $    3.7 \pm\ph{0}    0.2$ 	& $   15 \pm\ph{0}    2$ \\
4	& $B^+ \rar \pi^+\pi^0$ 						& $    5.5 \pm\ph{0}    0.6$ 	& $    14 \pm\ph{0}    2$ \\
4	& $B^+ \rar K^+\pi^0$ 							& $   12.1 \pm\ph{0}    0.8$ 	& $    8 \pm\ph{0}    1$ \\
5	& $B^+ \rar (K^{(**)}(1430) \pi)^+ \rar  (\Kp\pi\pi)^+$ 		& $   29.0 \pm\ph{0}    5.4$ 	& $    38 \pm\ph{0}    5$ \\
\hline
6	& $B^0 \rar \pi^- K^{\star +}(\rar K^0_S\pi^+)$ 			& $    3.3 \pm\ph{0}    0.4$ 	& $    2 \pm\ph{0}    1$ \\
7	& $B^0 \rar \rho^+\rho^-_{\rm\:[long]}$ 				& $   25.2 \pm\ph{0}   3.7$ 	& $   67 \pm\ph{}   10$ \\
7	& $B^0 \rar (a_1\pi)^0$ 						& $   39.7 \pm   3.7$ 		& $    39 \pm\ph{}    4$ \\
8	& $B^0 \rar K^+\pi^-$ 							& $   18.9 \pm\ph{0}    0.7$ 	& $    12 \pm\ph{0}    0$ \\
9	& $B^0 \rar \pi^- K^{\star +}(\rar K^+\pi^0)$ 				& $    3.3 \pm\ph{0}    0.4$ 	& $   20 \pm\ph{0}   2$ \\
9	& $B^0 \rar K^{(**)}(1430) \pi \rar K\pi\pi^0$ 				& $   11.2 \pm\ph{0}    2.2$ 	& $    212 \pm\ph{}    34$ \\
10	& $B^0\rar\gamma K^{\star 0}(892,1430)(\rar (K^+\pi^-)^0)$ 		& $   27.4 \pm\ph{0}    1.5$ 	& $    14 \pm\ph{0}    1$ \\
10	& $B^0 \rar \pi^0 K^{\star 0}(\rar K^+\pi^-)$ 				& $    1.3 \pm\ph{0}    0.5$ 	& $   9 \pm\ph{}    4$ \\
10	& $B^0 \rar \eta^\prime(\rar\rho^0\gamma)\pi^0$ 			& $    0.4 \pm\ph{0}    0.2$ 	& $    3 \pm\ph{0}    2$ \\
11	& $B^0 \rar \rho^- K^+$ 						& $    9.9 \pm\ph{0}    1.6$ 	& $   103 \pm\ph{}    17$ \\
12	& $B^0 \rar K^+\pi^-\pi^0_{\rm~[nonres]}$ 				& $    4.6 \pm\ph{0}    4.6$ 	& $   38 \pm\ph{0}    38$ \\
13	& $B^0 \rar \pi^0K^0_S(\rar\pi^+\pi^-)$ 				& $   5.8 \pm\ph{0}    0.5$ 	& $   50 \pm\ph{0}    4$ \\
14	& $B^0 \rar D^-(\rar\pi^-\pi^0)\pi^+$					& $    7.5 \pm\ph{0}   2.3$ 	& $  599 \pm\ph{}   184$ \\
15	& $B^0 \rar \Dzb(\rar K^+\pi^-)\pi^0$ 					& $   11.0 \pm\ph{0}    3.2$ 	& $   100 \pm\ph{0}    29$ \\
16	& $B^0 \rar \Dzb(\rar \pi^+\pi^-)\pi^0$ 				& $    0.4 \pm\ph{0}    0.1$ 	& $    35 \pm\ph{0}    9$ \\
17	& $B^0 \rar J/\psi(\rar e^+e^-,\mu^+\mu^-)\pi^0$ 			& $    2.6 \pm\ph{0}    0.5$ 	& $   77 \pm\ph{0}    15$ \\
\hline
18	& $B^0 \rar \{\text{neutral generic } b\rar c \text{ decays}\}$ 	& $    -$ 			& $173\pm\ph{0} 15$ \\
19	& $B^+ \rar \{\text{charged generic } b\rar c \text{ decays}\}$ 	& $    -$ 			& $396\pm\ph{} 20$ \\
\hline
\end{tabular*}
\vspace{-0.2cm}
\caption{ \label{tab:bbackground}
	Summary of the \B-background modes taken into account for the
	likelihood model. They have been grouped in twenty classes:
	charged charmless (six), neutral charmless (eight), 
	exclusive neutral charmed (four) and inclusive neutral and
	charged charmed decays. Modes with at least two events expected
	after final selection have been included.}
\end{center}
\end{table*}

We use MC simulated events to study the background from other $B$ 
decays. More than a hundred channels have been considered in the
preliminary studies, of which twenty-nine have been finally included
in the likelihood model -- decays with at least two events expected after
selection. These exclusive \B-background modes are grouped into eighteen 
different classes gathering decays with similar kinematic and topological
properties: six for charged charmless \B-decays, eight for neutral charmless
\B-decays and four for exclusive neutral charmed \B-decays.
Two additional classes account for inclusive neutral and charged 
$b\to c$ decays.

Table \ref{tab:bbackground} summarizes the twenty background classes which are
used in the fit. For each mode, the expected number of selected events is
computed by multiplying the selection efficiency (estimated using MC
simulated decays) by the branching fraction scaled up to the dataset 
luminosity ($310\;\mathrm{fb}^{-1}$). The world average branching ratios have been
used for the experimentally known decay modes. When only upper limits are
given, they have been translated into branching ratios using all information
available such as additional conservative hypotheses (e.g. 100\% longitudinal
polarization for $B\to\rho\rho$ decay) if needed.

\subsection{THE MAXIMUM-LIKELIHOOD FIT}
\label{subsec:ML}

We perform an unbinned extended maximum-likelihood fit to extract
the inclusive $\Btopipipi$ event yield and the $U$ and $I$ coefficients
defined in Eqs.~(\ref{eq:firstObs})--(\ref{eq:lastObs}). 
The fit uses the variables $\mes$, $\deprime$, the NN output, and the 
Dalitz plot to discriminate signal from background. The 
$\dt$ measurement allows to the determination of mixing-induced \CP violation
and provides additional continuum-background rejection. 

The selected on-resonance data sample is assumed to consist of signal, 
continuum-background and \B-background components, separated by the 
flavor and tagging category of the tag side \B decay. 
The signal likelihood consists of the sum of a correctly 
reconstructed (``truth-matched'', TM) component and a misreconstructed 
(``self-cross-feed'', SCF) component.

The probability density function (PDF) ${\cal P}_i^\cat$ for an
event $i$ in tagging category $\cat$ is the sum of the probability densities 
of all components, namely
\beqn
\label{eq:theLikelihood}
	{\cal P}_i^\cat
	&\equiv& 
		N_{\tpi} f^\cat_{\tpi}
		\left[ 	(1-\fscfave^\cat){\cal P}_{\tpi-\TM,i}^\cat +
			\fscfave^\cat{\cal P}_{\tpi-\SCF,i}^\cat 
		\right] 
		\nonumber\\[0.3cm]
	&&
		+\; N^\cat_{q\bar q}\frac{1}{2}
		\left(1 + \Qtagi\Atagqq\right){\cal P}_{q\bar q,i}^\cat
		\nonumber \\[0.3cm]
	&&
		+\; \sum_{j=1}^{N^{B^+}_{\rm class}}
		N_{B^+j} f^\cat_{B^+j}
		\frac{1}{2}\left(1 + \Qtagi \Atagj\right){\cal P}_{B^+,ij}^\cat
		\nonumber \\[0.3cm]
	&&
		+\; \sum_{j=1}^{N^{B^0}_{\rm class}}
		N_{B^0j} f^\cat_{B^0j}
		{\cal P}_{B^0,ij}^\cat~,
\eeqn
where: 
	$N_{\tpi}$ is the total number of $\pi^+\pi^-\pi^0$ signal events 
	in the data sample;
 	$f^\cat_{\tpi}$ is the fraction of signal events that are 
       	tagged in category $\cat$;
	$\fscfave^\cat$ is the fraction of SCF events in tagging category $\cat$, 
	averaged over the Dalitz plot;
	${\cal P}_{\tpi-\TM,i}^\cat$ and ${\cal P}_{\tpi-\SCF,i}^\cat$
	are the products of PDFs of the discriminating variables used
	in tagging category $\cat$ for TM and SCF
	events, respectively; 
 	$N^\cat_{q\bar q}$ is the number of continuum events that are 
	tagged in category $\cat$;
	$\Qtagi$ is the tag flavor of the event, defined to be 
	$+1$ for a $\Bz_{\rm tag}$ and $-1$ for a $\Bzb_{\rm tag}$; 
	$\Atagqq$ parameterizes possible tag asymmetry in continuum events; 
	${\cal P}_{q\bar q,i}^\cat$ is the continuum PDF for tagging 
	category $\cat$;
	$N^{B^+}_{\rm class}$ ($N^{B^0}_{\rm class}$) is the number of 
	charged (neutral) $B$-related background classes considered in the fit;
	$N_{B^+j}$ ($N_{B^0j}$) is the number of expected events in
	the charged (neutral) $B$-background class $j$;
	$f^\cat_{B^+j}$ ($f^\cat_{B^0j}$) is the fraction of 
	charged (neutral) $B$-background events of class $j$
	that are tagged in category $\cat$;
	$\Atagj$ describes a possible tag asymmetry in the charged-$B$ background
	class $j$; 
	correlations between the tag and the position in the Dalitz plot 
	(the ``charge'') are absorbed in tag-flavor-dependent 
	Dalitz plot PDFs that are used for charged-\B and continuum
	background;
	${\cal P}_{B^+,ij}^\cat$ is the $B^+$-background PDF for tagging 
	category $\cat$ and class $j$;
	finally, ${\cal P}_{B^0,ij}^\cat$ is the neutral-$B$-background 
	PDF for tagging category $\cat$ and class $j$.

The PDFs ${\cal P}_{X}^{\cat}$ ($X=\{TM, SCF, {\rm continuum}, {\rm B-bkg}\}$)
are the product of the four PDFs of the discriminating variables,
$x_1 = m_{ES}$, $x_2 = \deprime$, $x_3 = {\rm NN output}$, and the triplet
$x_4 = \{\mprime, \thetaprime, \deltat\}$:
\beq
\label{eq:likVars}
	{\cal P}_{X,i(j)}^{\cat} \;\equiv\; 
	\prod_{k=1}^4 P_{X,i(j)}^\cat(x_k)~.
\eeq
The extended likelihood over all tagging categories is given by
\beq
	{\cal L} \;\equiv\;  
	\prod_{\cat=1}^{7} e^{-\overline N^\cat}\,
	\prod_{i}^{N^\cat} {\cal P}_{i}^\cat~,
\eeq
where $\overline N^\cat$ is the total number of events expected in category 
$\cat$. 

A total of 68 parameters, including the inclusive signal yield and the
parameters from Eq.~(\ref{eq:dt}), are varied in the fit. Most of the
parameters describing the continuum distributions are also floated
in the fit.

\begin{table*}[t]
\begin{center}
\setlength{\tabcolsep}{0.0pc}
\begin{tabular*}{\textwidth}{@{\extracolsep{\fill}}l|cccc}
\hline
 Variable  	& TM Signal	&  SCF Signal  	&   Continuum  	&   B-Background  \\
\hline\\[-0.3cm]
$\de$ 		& 	GG	& G		& P2		&	NP	  \\
$\mes$ 		& 	biCB	& NP		& Argus		&	NP	  \\
Neural Net 	& 	NP	& NP		& P3		&	NP	  \\
Dalitz 		& see text	& see text	& NP		&	NP	  \\
$\dt$ 		& 	GGG	& GGG		& GGG		&	GGG	  \\
\hline
\end{tabular*}

\vspace{-0.2cm}
\caption{ \label{tab:pdfparameterization}
        Summary of PDF parameterizations where G=Gaussian, PX=X-order polynomial, NP=non-parametric, and biCB=bifurcated Crystal Ball. See Section \ref{sec:deltaT} for a detailed description of the Dalitz plot parameterization for signal.}
\end{center}
\end{table*}

\subsubsection{\boldmath THE $\dt$ AND DALITZ PLOT PDFS}
\label{sec:deltaT}

	The Dalitz plot PDFs require as input the Dalitz plot-dependent 
	relative selection efficiency, $\e=\e(\mprime,\thetaprime)$, 
	and SCF fraction, $\fscf=\fscf(\mprime,\thetaprime)$.
	Both quantities are taken from MC simulation. 
	Away from the Dalitz plot corners the efficiency is uniform, while it 
	decreases when approaching the corners, where one out of the 
	three bodies in the final state is close to rest so that the 
	acceptance requirements on the particle reconstruction become 
        restrictive.
	Combinatorial backgrounds and hence SCF fractions are large in
	the corners
	of the Dalitz plot due to the presence of soft neutral clusters 
	and tracks. 

	For an event~$i$, we define the time-dependent Dalitz plot PDFs
	\beqn
		P_{\tpi-\TM,i} &=&
		\varepsilon_i\,(1 - \fscfi)\,\detJi\,\AmpAll~,
		\\[0.3cm]
		P_{\tpi-\SCF,\,i} &=&
		\varepsilon_i\,\fscfi\,\detJi\,\AmpAll~,
	\eeqn	
	where $P_{\tpi-\TM,i}$ and $P_{\tpi-\SCF,\,i}$ are normalized. The 
	corresponding phase space integration involves the expectation values 	
	$\langle \varepsilon\,(1-\fscf)\,\detJ \,f^\kappa f^{\sigma*}\rangle$
	and 
	$\langle \varepsilon\,\fscf\,\detJ\, f^\kappa f^{\sigma*}\rangle$
	for TM and SCF events, where the indices $\kappa$, $\sigma$ 
	run over all resonances belonging to the signal model.
	The expectation values are model-dependent and are 
	computed with the use of MC integration over the square Dalitz plot:
	\beq
	\label{eq:normAverage}
		\langle \varepsilon\,(1-\fscf)\,\detJ\, f^\kappa f^{\sigma*}\rangle
		\;=\; \frac{\int_0^1\int_0^1 
			    \varepsilon\,(1-\fscf)\,\detJ\, f^\kappa f^{\sigma*}
			\,d\mprime d\thetaprime}
		       {\int_0^1\int_0^1 \varepsilon\,\detJ\, f^\kappa f^{\sigma*}
			\,d\mprime d\thetaprime}~,
	\eeq
	and similarly for 
	$\langle \varepsilon\,\,\detJ\, f^\kappa f^{\sigma*}\rangle$,
	where all quantities in the integrands are Dalitz plot-dependent.

	Equation~(\ref{eq:theLikelihood}) invokes the phase 
	space-averaged SCF fraction 
	$\fscfave\equiv\langle\fscf\,\detJ\, f^\kappa f^{\sigma*}\rangle$. 
	The PDF normalization  is decay-dynamics-dependent
	and is computed iteratively. We 
	determine the average SCF fractions separately for each tagging category 
	from MC simulation. 
	
	The width of the dominant $\rho(770)$ resonance is large compared 
	to the mass resolution for TM events (about $8\mevcc$ core Gaussian
	resolution). We  therefore neglect resolution effects in the TM 
	model.	
	Misreconstructed events	have a poor mass resolution that strongly 
	varies across the Dalitz plot. It is described in the fit by a 
	$2\times 2$-dimensional resolution function
	\beq
	\label{eq:rscf}
		\Rscf(\mprime_r,\thetaprime_r,\mprime_t,\thetaprime_t)~,
	\eeq
	which represents the probability to reconstruct at the coordinate
	$(\mprime_r,\thetaprime_r)$ an event that has the true coordinate 
	$(\mprime_t,\thetaprime_t)$. It obeys the unitarity condition
	\beq
		\intl_0^1\intl_0^1 
		\Rscf(\mprime_r,\thetaprime_r,\mprime_t,\thetaprime_t)
		\,d\mprime_r d\thetaprime_r = 1,~
	\eeq
	and is convolved with the signal model. 
	The $\Rscf$ function is obtained from MC simulation.

	We use the signal model described in Section~\ref{sec:kinmeatics}. 
	It contains the dynamical information and is connected with $\dt$ via 
	the matrix element~(\ref{eq:dt}), which serves as PDF. It is diluted 
	by the effects of mistagging and the limited vertex 
	resolution~\cite{rhopipaper}. 
	The $\deltat$ resolution function for signal and \B-background 
	events is a sum of three Gaussian distributions, with parameters 
	determined by a fit to fully reconstructed $\Bz$ 
	decays~\cite{BabarS2b}. 
\\[0.3cm]\noindent
	The Dalitz plot- and $\dt$-dependent PDFs factorize for the 
	charged-$B$-background modes, but not (necessarily) 
	for the neutral-$B$ background due to $\BzBzb$ mixing.

 	The charged \B-background
		contribution to the likelihood~(\ref{eq:theLikelihood})
                involves 
		the parameter $\Atag$, multiplied by the tag flavor $\Qtag$ of 
		the event. In the presence of significant tag-``charge'' 
		correlation (represented by an effective 
		flavor-tag-versus-Dalitz-coordinate correlation),
		it parameterizes possible direct \CP violation in these events.
		We also use distinct square Dalitz plot PDFs for each 
		reconstructed $B$ flavor tag, and a flavor-tag-averaged PDF for 
		untagged events. The PDFs are obtained from MC simulation and are 
		described with the use of non-parametric functions.
		The $\dt$ resolution parameters are determined by a fit to fully 
		reconstructed $\Bp$ decays. For each $\Bp$-background class we adjust 
		effective lifetimes to account for the misreconstruction of the 
		event that modifies the nominal $\dt$ resolution function.

	The neutral-$B$ background is parameterized with PDFs that
		depend on the flavor tag of the event. In the case of \CP
		eigenstates, correlations between the flavor tag and the Dalitz 
		coordinate are expected to be small. However, non-\CP  eigenstates,
		such as $a_1^\pm\pi^\mp$, may exhibit such correlation. Both types 
		of decays can have direct
		and mixing-induced \CP  violation. A third type of decays
		involves charged kaons and does not exhibit mixing-induced
		\CP  violation, but usually has a strong correlation between the
		flavor tag and the Dalitz plot coordinate (the kaon charge), because 
		it consists of $B$-flavor eigenstates.
		The Dalitz plot PDFs are obtained from MC simulation and are 
		described with the use of non-parametric functions.
		For neutral $B$ background, the signal $\dt$ resolution model 
		is assumed.

	The Dalitz plot
		treatment of the continuum events is similar to the one used
		for charged-$B$ background. 
		The square Dalitz plot PDF for continuum background is 
		obtained from on-resonance events selected in the
		$\mes$ sidebands and corrected for feed-through
		from \B decays. A large number of cross checks has been 
		performed to ensure the high fidelity of the empirical shape 
		parameterization. Analytical models have been found insufficient.
		The continuum $\deltat$ distribution is parameterized as the sum of 
		three Gaussian distributions with common mean and 
		three distinct widths that scale the $\dt$ per-event error. 
		This yields six shape parameters that are determined by 
		the fit.
 		The model is motivated by the observation that 
		the $\dt$ average is independent of its error, and that the 
		$\dt$ RMS depends linearly on the $\dt$ error.

\subsubsection{PARAMETERIZATION OF THE OTHER VARIABLES}
\label{sec:likmESanddE}

	The $\mes$ distribution of TM signal events is
		parameterized by a bifurcated Crystal Ball function~\cite{PDFsCB},
		which is a combination of a one-sided Gaussian and 
		a Crystal Ball function. The mean of this function
		is determined by the fit. A non-parametric
		function is used to describe the SCF signal component.

	The $\deprime$ distribution of TM events is
		parameterized by a double Gaussian function, where
		all five parameters depend linearly on $\mpm^2$.
		Misreconstructed events are parameterized by a broad
		single Gaussian function.
		
	Both $\mes$ and $\deprime$ PDFs are parameterized by non-parametric
		functions for all $B$-background classes.

	The $\mes$ and $\deprime$ PDFs for continuum events are
		parameterized with an Argus shape function~\cite{PDFsArgus} and 
		a second order polynomial, respectively, with parameters 
		determined by the fit.

	We use non-parametric functions to empirically describe the distributions 
		of the NN outputs
		found in the MC simulation for TM and SCF signal events, 
		and for \B-background events. We distinguish tagging categories 
		for TM signal events to account for differences observed in the 
		shapes.
	
	The continuum NN distribution is parameterized by a 
		third order polynomial that is defined to be positive. 
		The coefficients of the polynomial are determined by the fit.
		Continuum events exhibit a correlation between the Dalitz plot 
		coordinate
		and the shape of the event that is exploited in the NN. 
To correct for residual effects,
		we introduce a linear dependence of the polynomial coefficients
		on the distance of the Dalitz plot coordinate to the kinematic 
		boundaries of the Dalitz plot. The parameters describing this
		dependence are determined by the fit.

\section{SYSTEMATIC STUDIES}
\label{sec:Systematics}

\begin{table*}[t]
\begin{center}
\setlength{\tabcolsep}{0.0pc}
\begin{tabular*}{\textwidth}{@{\extracolsep{\fill}}lcccccc}
\hline
 \rule[-6pt]{0pt}{18pt}  &      $\Iz$  &      $\IM$  &   $\IMzIm$  &   $\IMzRe$  &      $\IP$  &   $\IPzIm$  \\
\hline\\[-0.35cm]
 Dalitz plot model  &  0.010  &  0.006  &  0.110  &  0.102  &  0.020  &  0.018  \\
 $\rho$,$\rho'$,$\rho''$ lineshape   &  0.003  &  0.012  &  0.240  &  0.103  &  0.009  &  0.225  \\
   Fit bias  &  0.014  &  0.023  &  0.173  &  0.375  &  0.008  &  0.186  \\
 $N_{\rm Background}$  &  0.002  &  0.008  &  0.064  &  0.085  &  0.005  &  0.026  \\
    \B background \CP  &  0.005  &  0.009  &  0.082  &  0.061  &  0.011  &  0.039  \\
               Others  &  0.002  &  0.006  &  0.104  &  0.052  &  0.005  &  0.051  \\
\hline\\[-0.3cm]
                  Sum  &  0.017  &  0.027  &  0.315  &  0.402  &  0.023  &  0.292  \\
\hline
\end{tabular*}
\vspace{1.5\baselineskip}

\vspace{-0.35cm}

\setlength{\tabcolsep}{0.0pc}
\begin{tabular*}{\textwidth}{@{\extracolsep{\fill}}lccccccc}
\hline
 \rule[-6pt]{0pt}{18pt}  &    $\IPzRe$  &    $\IPMIm$  &    $\IPMRe$  &      $\Uzm$  &      $\Uzp$  &   $\UMzmIm$  &   $\UMzmRe$  \\
\hline\\[-0.3cm]
 Dalitz plot model  &  0.017  &  0.007  &  0.127  &  0.082  &  0.041  &  0.144  &  0.209  \\
 $\rho$,$\rho'$,$\rho''$ lineshape  &  0.308  &  0.138  &  0.306  &  0.012  &  0.012  &  0.086  &  0.159  \\
   Fit bias  &  0.301  &  0.036  &  0.048  &  0.088  &  0.001  &  0.050  &  0.087  \\
 $N_{\rm Background}$  &  0.049  &  0.178  &  0.176  &  0.002  &  0.009  &  0.034  &  0.045  \\
    \B background \CP  &  0.042  &  0.044  &  0.095  &  0.015  &  0.002  &  0.020  &  0.073  \\
               Others  &  0.073  &  0.059  &  0.095  &  0.004  &  0.002  &  0.012  &  0.050  \\
\hline\\[-0.3cm]
                  Sum  &  0.431  &  0.143  &  0.335  &  0.121  &  0.043  &  0.175  &  0.276  \\
\hline
\end{tabular*}
\vspace{1.5\baselineskip}

\vspace{-0.35cm}
\setlength{\tabcolsep}{0.0pc}
\begin{tabular*}{\textwidth}{@{\extracolsep{\fill}}lccccccc}
\hline
 \rule[-6pt]{0pt}{18pt}  &   $\UMzpIm$  &   $\UMzpRe$  &      $\UMm$  &      $\UMp$  &   $\UPzmIm$  &   $\UPzmRe$  &   $\UPzpIm$  \\
\hline\\[-0.3cm]
 Dalitz plot model  &  0.034  &  0.024  &  0.022  &  0.030  &  0.036  &  0.258  &  0.076  \\
 $\rho$,$\rho'$,$\rho''$ lineshape  &  0.222  &  0.045  &  0.010  &  0.030  &  0.050  &  0.216  &  0.089  \\
   Fit bias  &  0.034  &  0.058  &  0.004  &  0.005  &  0.053  &  0.007  &  0.004  \\
 $N_{\rm Background}$  &  0.045  &  0.017  &  0.013  &  0.005  &  0.117  &  0.103  &  0.034  \\
    \B background \CP  &  0.032  &  0.018  &  0.032  &  0.009  &  0.063  &  0.075  &  0.019  \\
               Others  &  0.027  &  0.015  &  0.010  &  0.002  &  0.045  &  0.020  &  0.027  \\
\hline\\[-0.3cm]
                  Sum  &  0.227  &  0.078  &  0.025  &  0.043  &  0.082  &  0.337  &  0.117  \\
\hline
\end{tabular*}
\vspace{1.5\baselineskip}

\vspace{-0.35cm}
\setlength{\tabcolsep}{0.0pc}
\begin{tabular*}{\textwidth}{@{\extracolsep{\fill}}lcccccc}
\hline
 \rule[-6pt]{0pt}{18pt}  &   $\UPzpRe$  &   $\UPMmIm$  &   $\UPMmRe$  &   $\UPMpIm$  &   $\UPMpRe$  &      $\UPm$ \\
\hline\\[-0.3cm]
 Dalitz plot model  &  0.045  &  0.014  &  0.250  &  0.703  &  0.227  &  0.010 \\
 $\rho$,$\rho'$,$\rho''$ lineshape  &  0.140  &  0.169  &  0.200  &  0.169  &  0.159  &  0.031 \\
   Fit bias  &  0.020  &  0.007  &  0.069  &  0.012  &  0.033  &  0.027 \\
 $N_{\rm Background}$  &  0.069  &  0.137  &  0.122  &  0.024  &  0.166  &  0.014 \\
    \B background \CP  &  0.014  &  0.042  &  0.029  &  0.025  &  0.025  &  0.034 \\
               Others  &  0.024  &  0.032  &  0.067  &  0.024  &  0.044  &  0.009 \\
\hline\\[-0.3cm]
                  Sum  &  0.148  &  0.170  &  0.328  &  0.723  &  0.279  &  0.042 \\
\hline
\end{tabular*}
\vspace{1.5\baselineskip}

\vspace{-0.35cm}
\setlength{\tabcolsep}{0.0pc}
\begin{tabular*}{\textwidth}{@{\extracolsep{\fill}}lccccc}
\hline
\rule[-6pt]{0pt}{18pt}  &   $\Acp$  &    $C$  &    $\dC$  &     $S$  &    $\dS$  \\
\hline\\[-0.3cm]
 Dalitz plot model  &  0.008  &  0.002  &  0.013  &  0.015  &  0.024  \\
 $\rho$,$\rho'$,$\rho''$ lineshape  &  0.011  &  0.021  &  0.011  &  0.016  &  0.008  \\
   Fit bias  &  0.004  &  0.012  &  0.015  &  0.028  &  0.011  \\
 $N_{\rm Background}$  &   0.002  &   0.005  &   0.011  &   0.002  &   0.011   \\
    \B background \CP  &   0.003  &   0.029  &   0.006  &   0.017  &   0.005   \\
               Others  &   0.001  &   0.007  &   0.005  &   0.006  &   0.007   \\
\hline\\[-0.3cm]
                  Sum  &   0.015  &   0.037  &   0.021  &   0.028  &   0.029   \\
\hline
\end{tabular*}

\vspace{-0.2cm}
\caption{ \label{tab:systematics}
        Summary of systematic uncertainties.}
\end{center}
\end{table*}

\begin{table*}[t]
\centering
\begin{small}
\setlength{\tabcolsep}{0.0pc}
\begin{tabular*}{\textwidth}{@{\extracolsep{\fill}}lccccccccccccc}
\hline
 && \\[-0.3cm]
\rule[-6pt]{0pt}{18pt}  & $\sigPiNb$  &      $\Iz$  &      $\IM$  &   $\IMzIm$  &   $\IMzRe$  &      $\IP$  &   $\IPzIm$  &   $\IPzRe$  &   $\IPMIm$  &   $\IPMRe$  &     $\Uzm$  &     $\Uzp$  &  $\UMzmIm$ \\[0.15cm]
\hline
 &&\\[-0.3cm] $\sigPiNb$  &  $\phantom{-}1.00$  &  &  &  &  &  &  &  &  &  &  &  & \\
      $\Iz$  &  $\phantom{-}0.29$  &  $\phantom{-}1.00$  &  &  &  &  &  &  &  &  &  &  & \\
      $\IM$  &  $\phantom{-}0.12$  &  $-0.68$  &  $\phantom{-}1.00$  &  &  &  &  &  &  &  &  &  & \\
   $\IMzIm$  &  $-0.18$  &  $-0.61$  &  $\phantom{-}0.73$  &  $\phantom{-}1.00$  &  &  &  &  &  &  &  &  & \\
   $\IMzRe$  &  $-0.07$  &  $\phantom{-}0.84$  &  $-0.82$  &  $-0.60$  &  $\phantom{-}1.00$  &  &  &  &  &  &  &  & \\
      $\IP$  &  $\phantom{-}0.72$  &  $\phantom{-}0.18$  &  $\phantom{-}0.24$  &  $-0.03$  &  $-0.05$  &  $\phantom{-}1.00$  &  &  &  &  &  &  & \\
   $\IPzIm$  &  $\phantom{-}0.10$  &  $-0.57$  &  $\phantom{-}0.73$  &  $\phantom{-}0.75$  &  $-0.61$  &  $\phantom{-}0.40$  &  $\phantom{-}1.00$  &  &  &  &  &  & \\
   $\IPzRe$  &  $-0.28$  &  $\phantom{-}0.59$  &  $-0.66$  &  $-0.43$  &  $\phantom{-}0.62$  &  $-0.48$  &  $-0.79$  &  $\phantom{-}1.00$  &  &  &  &  & \\
   $\IPMIm$  &  $\phantom{-}0.18$  &  $\phantom{-}0.13$  &  $\phantom{-}0.14$  &  $\phantom{-}0.01$  &  $-0.08$  &  $\phantom{-}0.03$  &  $\phantom{-}0.23$  &  $-0.21$  &  $\phantom{-}1.00$  &  &  &  & \\
   $\IPMRe$  &  $\phantom{-}0.35$  &  $\phantom{-}0.12$  &  $\phantom{-}0.08$  &  $\phantom{-}0.13$  &  $-0.17$  &  $-0.09$  &  $-0.14$  &  $\phantom{-}0.38$  &  $-0.01$  &  $\phantom{-}1.00$  &  &  & \\
     $\Uzm$  &  $\phantom{-}0.79$  &  $-0.15$  &  $\phantom{-}0.41$  &  $\phantom{-}0.13$  &  $-0.49$  &  $\phantom{-}0.70$  &  $\phantom{-}0.38$  &  $-0.46$  &  $-0.10$  &  $\phantom{-}0.36$  &  $\phantom{-}1.00$  &  & \\
     $\Uzp$  &  $-0.83$  &  $-0.53$  &  $\phantom{-}0.27$  &  $\phantom{-}0.46$  &  $-0.22$  &  $-0.69$  &  $\phantom{-}0.16$  &  $\phantom{-}0.01$  &  $-0.00$  &  $-0.22$  &  $-0.66$  &  $\phantom{-}1.00$  & \\
  $\UMzmIm$  &  $-0.83$  &  $-0.28$  &  $\phantom{-}0.01$  &  $\phantom{-}0.26$  &  $\phantom{-}0.07$  &  $-0.47$  &  $\phantom{-}0.17$  &  $-0.08$  &  $\phantom{-}0.13$  &  $-0.65$  &  $-0.77$  &  $\phantom{-}0.79$  &  $\phantom{-}1.00$ \\
  $\UMzmRe$  &  $\phantom{-}0.70$  &  $\phantom{-}0.23$  &  $-0.05$  &  $-0.33$  &  $-0.13$  &  $\phantom{-}0.51$  &  $-0.07$  &  $\phantom{-}0.05$  &  $\phantom{-}0.01$  &  $\phantom{-}0.51$  &  $\phantom{-}0.72$  &  $-0.76$  &  $-0.83$ \\
  $\UMzpIm$  &  $-0.22$  &  $\phantom{-}0.21$  &  $-0.41$  &  $-0.57$  &  $\phantom{-}0.10$  &  $-0.29$  &  $-0.52$  &  $\phantom{-}0.47$  &  $-0.00$  &  $\phantom{-}0.03$  &  $-0.12$  &  $-0.09$  &  $-0.05$ \\
  $\UMzpRe$  &  $-0.39$  &  $\phantom{-}0.32$  &  $-0.43$  &  $-0.27$  &  $\phantom{-}0.55$  &  $-0.31$  &  $-0.24$  &  $\phantom{-}0.14$  &  $\phantom{-}0.31$  &  $-0.51$  &  $-0.76$  &  $\phantom{-}0.27$  &  $\phantom{-}0.60$ \\
     $\UMm$  &  $-0.36$  &  $-0.25$  &  $\phantom{-}0.12$  &  $\phantom{-}0.05$  &  $-0.22$  &  $-0.39$  &  $-0.06$  &  $\phantom{-}0.04$  &  $\phantom{-}0.05$  &  $-0.08$  &  $-0.22$  &  $\phantom{-}0.40$  &  $\phantom{-}0.30$ \\
     $\UMp$  &  $\phantom{-}0.48$  &  $\phantom{-}0.57$  &  $-0.40$  &  $-0.45$  &  $\phantom{-}0.24$  &  $\phantom{-}0.36$  &  $-0.32$  &  $\phantom{-}0.42$  &  $-0.05$  &  $\phantom{-}0.47$  &  $\phantom{-}0.44$  &  $-0.73$  &  $-0.68$ \\
  $\UPzmIm$  &  $\phantom{-}0.30$  &  $\phantom{-}0.42$  &  $-0.24$  &  $-0.51$  &  $\phantom{-}0.28$  &  $\phantom{-}0.04$  &  $-0.25$  &  $\phantom{-}0.11$  &  $\phantom{-}0.55$  &  $-0.03$  &  $-0.09$  &  $-0.27$  &  $-0.15$ \\
  $\UPzmRe$  &  $-0.61$  &  $-0.29$  &  $-0.01$  &  $-0.08$  &  $-0.24$  &  $-0.76$  &  $-0.34$  &  $\phantom{-}0.38$  &  $\phantom{-}0.03$  &  $\phantom{-}0.12$  &  $-0.45$  &  $\phantom{-}0.57$  &  $\phantom{-}0.31$ \\
  $\UPzpIm$  &  $-0.57$  &  $-0.37$  &  $\phantom{-}0.04$  &  $\phantom{-}0.01$  &  $-0.06$  &  $-0.18$  &  $\phantom{-}0.22$  &  $-0.37$  &  $\phantom{-}0.01$  &  $-0.84$  &  $-0.40$  &  $\phantom{-}0.47$  &  $\phantom{-}0.77$ \\
  $\UPzpRe$  &  $\phantom{-}0.40$  &  $\phantom{-}0.15$  &  $\phantom{-}0.10$  &  $-0.02$  &  $\phantom{-}0.15$  &  $\phantom{-}0.34$  &  $\phantom{-}0.25$  &  $-0.50$  &  $\phantom{-}0.45$  &  $-0.35$  &  $\phantom{-}0.02$  &  $-0.18$  &  $\phantom{-}0.04$ \\
  $\UPMmIm$  &  $-0.17$  &  $-0.20$  &  $\phantom{-}0.22$  &  $\phantom{-}0.61$  &  $\phantom{-}0.02$  &  $\phantom{-}0.20$  &  $\phantom{-}0.57$  &  $-0.35$  &  $-0.15$  &  $-0.28$  &  $-0.07$  &  $\phantom{-}0.22$  &  $\phantom{-}0.40$ \\
  $\UPMmRe$  &  $\phantom{-}0.45$  &  $\phantom{-}0.53$  &  $-0.25$  &  $-0.01$  &  $\phantom{-}0.36$  &  $\phantom{-}0.45$  &  $-0.03$  &  $\phantom{-}0.33$  &  $-0.20$  &  $\phantom{-}0.48$  &  $\phantom{-}0.35$  &  $-0.56$  &  $-0.53$ \\
  $\UPMpIm$  &  $-0.89$  &  $-0.54$  &  $\phantom{-}0.19$  &  $\phantom{-}0.31$  &  $-0.28$  &  $-0.68$  &  $\phantom{-}0.12$  &  $-0.04$  &  $\phantom{-}0.06$  &  $-0.38$  &  $-0.64$  &  $\phantom{-}0.88$  &  $\phantom{-}0.84$ \\
  $\UPMpRe$  &  $\phantom{-}0.56$  &  $\phantom{-}0.20$  &  $\phantom{-}0.00$  &  $\phantom{-}0.13$  &  $\phantom{-}0.18$  &  $\phantom{-}0.70$  &  $\phantom{-}0.28$  &  $-0.30$  &  $-0.29$  &  $\phantom{-}0.00$  &  $\phantom{-}0.50$  &  $-0.55$  &  $-0.37$ \\
     $\UPm$  &  $-0.04$  &  $\phantom{-}0.56$  &  $-0.54$  &  $-0.44$  &  $\phantom{-}0.44$  &  $-0.16$  &  $-0.53$  &  $\phantom{-}0.65$  &  $-0.02$  &  $\phantom{-}0.30$  &  $-0.13$  &  $-0.23$  &  $-0.18$ \\
[0.15cm]\hline
\end{tabular*}

\vspace{1.5\baselineskip}
\setlength{\tabcolsep}{0.0pc}
\begin{tabular*}{\textwidth}{@{\extracolsep{\fill}}lcccccccccccccc}
\hline
 && \\[-0.3cm]
\rule[-6pt]{0pt}{18pt}  &  $\UMzmRe$  &  $\UMzpIm$  &  $\UMzpRe$  &     $\UMm$  &     $\UMp$  &  $\UPzmIm$  &  $\UPzmRe$  &  $\UPzpIm$  &  $\UPzpRe$  &  $\UPMmIm$  &  $\UPMmRe$  &  $\UPMpIm$  &  $\UPMpRe$  &     $\UPm$ \\[0.15cm]
\hline
 &&\\[-0.3cm]  $\UMzmRe$  &  $\phantom{-}1.00$  &  &  &  &  &  &  &  &  &  &  &  &  & \\
  $\UMzpIm$  &  $\phantom{-}0.33$  &  $\phantom{-}1.00$  &  &  &  &  &  &  &  &  &  &  &  & \\
  $\UMzpRe$  &  $-0.54$  &  $-0.07$  &  $\phantom{-}1.00$  &  &  &  &  &  &  &  &  &  &  & \\
     $\UMm$  &  $-0.34$  &  $\phantom{-}0.22$  &  $\phantom{-}0.02$  &  $\phantom{-}1.00$  &  &  &  &  &  &  &  &  &  & \\
     $\UMp$  &  $\phantom{-}0.80$  &  $\phantom{-}0.50$  &  $-0.41$  &  $-0.31$  &  $\phantom{-}1.00$  &  &  &  &  &  &  &  &  & \\
  $\UPzmIm$  &  $\phantom{-}0.22$  &  $\phantom{-}0.21$  &  $\phantom{-}0.32$  &  $\phantom{-}0.10$  &  $\phantom{-}0.18$  &  $\phantom{-}1.00$  &  &  &  &  &  &  &  & \\
  $\UPzmRe$  &  $-0.15$  &  $\phantom{-}0.60$  &  $\phantom{-}0.07$  &  $\phantom{-}0.49$  &  $-0.15$  &  $\phantom{-}0.11$  &  $\phantom{-}1.00$  &  &  &  &  &  &  & \\
  $\UPzpIm$  &  $-0.55$  &  $\phantom{-}0.17$  &  $\phantom{-}0.40$  &  $\phantom{-}0.27$  &  $-0.52$  &  $-0.10$  &  $\phantom{-}0.17$  &  $\phantom{-}1.00$  &  &  &  &  &  & \\
  $\UPzpRe$  &  $-0.20$  &  $-0.59$  &  $\phantom{-}0.48$  &  $-0.16$  &  $-0.37$  &  $\phantom{-}0.42$  &  $-0.54$  &  $\phantom{-}0.10$  &  $\phantom{-}1.00$  &  &  &  &  & \\
  $\UPMmIm$  &  $-0.48$  &  $-0.70$  &  $\phantom{-}0.10$  &  $-0.19$  &  $-0.39$  &  $-0.53$  &  $-0.56$  &  $\phantom{-}0.21$  &  $\phantom{-}0.23$  &  $\phantom{-}1.00$  &  &  &  & \\
  $\UPMmRe$  &  $\phantom{-}0.48$  &  $-0.20$  &  $-0.30$  &  $-0.51$  &  $\phantom{-}0.65$  &  $-0.08$  &  $-0.55$  &  $-0.69$  &  $-0.13$  &  $\phantom{-}0.21$  &  $\phantom{-}1.00$  &  &  & \\
  $\UPMpIm$  &  $-0.68$  &  $\phantom{-}0.21$  &  $\phantom{-}0.30$  &  $\phantom{-}0.48$  &  $-0.62$  &  $-0.24$  &  $\phantom{-}0.68$  &  $\phantom{-}0.68$  &  $-0.28$  &  $\phantom{-}0.09$  &  $-0.72$  &  $\phantom{-}1.00$  &  & \\
  $\UPMpRe$  &  $\phantom{-}0.20$  &  $-0.62$  &  $-0.19$  &  $-0.47$  &  $\phantom{-}0.18$  &  $-0.27$  &  $-0.91$  &  $-0.28$  &  $\phantom{-}0.35$  &  $\phantom{-}0.60$  &  $\phantom{-}0.61$  &  $-0.68$  &  $\phantom{-}1.00$  & \\
     $\UPm$  &  $\phantom{-}0.20$  &  $\phantom{-}0.48$  &  $-0.02$  &  $\phantom{-}0.34$  &  $\phantom{-}0.56$  &  $\phantom{-}0.30$  &  $\phantom{-}0.19$  &  $-0.29$  &  $-0.36$  &  $-0.33$  &  $\phantom{-}0.31$  &  $-0.17$  &  $-0.14$  &  $\phantom{-}1.00$ \\
[0.15cm]\hline
\end{tabular*}

\end{small}
\caption{\label{tab:corrmatSyst}
  Correlation matrix of systematic uncertainties for the $U$
  and $I$ coefficients.  Note that
  all elements above the diagonal are omitted for readability.
}
\label{tab:sysCorrelationsFinal}
\end{table*}

The contributions to the systematic error on the signal parameters are 
summarized in Table~\ref{tab:systematics}. Table~\ref{tab:corrmatSyst}
summarizes the correlation coefficient extracted from the
systematic covariance matrix.

The most important contribution to the systematic uncertainty stems
from the signal modeling of the Dalitz plot dynamics. 

To estimate the contribution to $\Btopipipi$ decay via other
resonances and non-resonant decays, we have
performed an independent analysis where we include these other decays in the fit model.
For simplicity, we assume a uniform Dalitz distribution
for the non-resonance events and consider possible resonances
including $f_0(980)$, $f_2(1270)$, and a low mass $s$-wave $\sigma$.
The fit does not find significant number of any of those decays.
However, the inclusion of a low mass $\pipi$ $s$-wave
component degrades our ability to identify $\rho^0\piz$ events 
significantly.
The systematic effects (contained in the ``Dalitz plot model'' field in 
Table~\ref{tab:systematics})
is estimated by observing the difference
between the true values and Monte Carlo fit results, in which
events are generated based on the new
fit results and fit with the nominal setup where only $\rho$ is taken
into account.

We vary the mass and width of the $\rho(770)$, $\rho(1450)$ and $\rho(1700)$ 
within ranges that 
exceed twice the errors found for these parameters in the fits to 
$\tau$ and $\epem$ data~\cite{taueeref}, and assign the observed
differences in the measured $U$ and $I$ coefficients as systematic uncertainties
(``$\rho,\rho',\rho''$ lineshape'' in Table~\ref{tab:systematics}). 
Since some of the $U$ and 
$I$ coefficients exhibit significant dependence on the $\rho(1450)$ 
and $\rho(1700)$ contributions, we leave their amplitudes (phases and 
fractions) free to vary in the nominal fit.

To validate the fitting tool, we perform fits on large MC samples with
the measured proportions of signal, continuum and $B$-background events.
No significant biases are observed in these fits. The statistical
uncertainties on the fit parameters are taken as systematic uncertainties
(``Fit bias'' in Table~\ref{tab:systematics}).

Another major source of systematic uncertainty is the $B$-background model. 
The expected event yields from the background modes are varied according 
to the uncertainties in the measured or estimated branching fractions
(``$N_{\rm{Background}}$'' in Table~\ref{tab:systematics}).  
Since $B$-background modes may exhibit  \CP violation, the corresponding 
parameters are varied within appropriate uncertainty ranges
(``$B$ background \CP'' in Table~\ref{tab:systematics}). As is done
for the signal PDFs, we vary the $\dt$ resolution parameters and
the flavor-tagging parameters within their uncertainties and assign
the differences observed in the on-resonance data fit with respect to
nominal fit as systematic errors.

Other systematic effects are much less important to the measurements
of $U$ and $I$ coefficients, and they are combined in 
``Others'' field in Table~\ref{tab:systematics}. Details are given
below.

The parameters for 
the continuum events are determined by the fit. No additional systematic 
uncertainties are assigned to them. An exception to this is the Dalitz 
plot PDF: to estimate the systematic
uncertainty from the \mes sideband extrapolation, we select large 
samples of off-resonance data by loosening the requirements on \de and 
the NN. We compare the distributions of $\mprime$ and $\thetaprime$ 
between the \mes sideband and the signal region. No significant 
differences are found. We assign as systematic error the effect seen when
weighting the continuum Dalitz plot PDF by the ratio of both data 
sets. This effect is mostly statistical in origin. 

The uncertainties associated with $\dmd$ and $\tau$ are
estimated by varying these parameters within the uncertainties
on the world average~\cite{PDG}.

The systematic effects due to the signal PDFs 
comprise uncertainties in the
PDF parameterization, the treatment of misreconstructed events, the
tagging performance, and the modeling of the signal contributions.

When the signal PDFs are determined from fits to a control sample
of fully reconstructed \B decays to exclusive final states with
charm, the uncertainties are obtained by varying the parameters
within the statistical uncertainties.
In other cases, the dominant parameters have been
left free to vary in the fit, and the differences observed in these
fits are taken as systematic errors.

The average fraction of misreconstructed signal events predicted by the MC
simulation has been verified with fully reconstructed $\B\to D\rho$
events~\cite{rhopipaper}. No significant differences between data and
the simulation were found. We vary $\fscfave$ for all tagging categories
relatively by $25\%$ to estimate the systematic uncertainty.

Tagging efficiencies, dilutions and biases for signal events
are varied within their experimental uncertainties.

The systematic errors for the parameters that measure interference
effects are dominated by the uncertainty in the signal model, mainly
the tail description of the $\rho$ resonance. For the other parameters,
the uncertainty on the fit bias and the \B-background contamination 
are important.

\section{FIT RESULTS}
\label{sec:fitResults}

\begin{table*}[t]
\begin{center}
\setlength{\tabcolsep}{0.3pc}
\begin{tabular*}{\textwidth}{@{\extracolsep{\fill}}llc}
\hline
&&\\[-0.3cm]
Parameter   & Description & Result \\[0.15cm]
\hline
&&\\[-0.3cm]
\rule[-1.7mm]{0mm}{5mm}$U_0^+$
     & Coefficient of $|f_0|^2$               & $\phantom{-}0.237\pm0.053\pm0.043$ \\
\rule[-1.7mm]{0mm}{5mm}$U_-^+$
     & Coefficient of $|f_-|^2$               & $\phantom{-}1.33\pm0.11\pm0.04$  \\[0.15cm]

\rule[-1.7mm]{0mm}{5mm}$U_0^-$
     & Coefficient of $|f_0|^2\cos(\dmd\dmt)$ & $-0.055\pm0.098\pm0.13$  \\
\rule[-1.7mm]{0mm}{5mm}$U_-^-$
     & Coefficient of $|f_-|^2\cos(\dmd\dmt)$ & $-0.30\pm0.15\pm0.03$  \\
\rule[-1.7mm]{0mm}{5mm}$U_+^-$
     & Coefficient of $|f_+|^2\cos(\dmd\dmt)$ & $\phantom{-}0.53\pm0.15\pm0.04$  \\[0.15cm]

\rule[-1.7mm]{0mm}{5mm}$I_0$
     & Coefficient of $|f_0|^2\sin(\dmd\dmt)$ & $-0.028\pm0.058\pm0.02$ \\
\rule[-1.7mm]{0mm}{5mm}$I_-$
     & Coefficient of $|f_-|^2\sin(\dmd\dmt)$ & $-0.03\pm0.10\pm0.03$ \\
\rule[-1.7mm]{0mm}{5mm}$I_+$
     & Coefficient of $|f_+|^2\sin(\dmd\dmt)$ & $\phantom{-}0.039\pm0.097\pm0.02$ \\[0.15cm]

\rule[-1.7mm]{0mm}{5mm}$U_{+-}^{+,\I}$
     & Coefficient of $\I[f_+f_-^*]$ & $\phantom{-}0.62\pm0.54\pm0.72$   \\
\rule[-1.7mm]{0mm}{5mm}$U_{+-}^{+,\R}$
     & Coefficient of $\R[f_+f_-^*]$ & $\phantom{-}0.38\pm0.55\pm0.28$    \\
\rule[-1.7mm]{0mm}{5mm}$U_{+-}^{-,\I}$
     & Coefficient of $\I[f_+f_-^*]\cos(\dmd\dmt)$ & $\phantom{-}0.13\pm0.94\pm0.17$  \\
\rule[-1.7mm]{0mm}{5mm}$U_{+-}^{-,\R}$
     & Coefficient of $\R[f_+f_-^*]\cos(\dmd\dmt)$ & $\phantom{-}2.14\pm0.91\pm0.33$   \\
\rule[-1.7mm]{0mm}{5mm}$I_{+-}^{\I}$
     & Coefficient of $\I[f_+f_-^*]\sin(\dmd\dmt)$ & $-1.9\ph{0}\pm1.1\ph{0}\pm0.1\ph{0}$  \\
\rule[-1.7mm]{0mm}{5mm}$I_{+-}^{\R}$
     & Coefficient of $\R[f_+f_-^*]\sin(\dmd\dmt)$ & $-0.1\ph{0}\pm1.9\ph{0}\pm0.3\ph{0}$\\[0.15cm]

\rule[-1.7mm]{0mm}{5mm}$U_{+0}^{+,\I}$
     & Coefficient of $\I[f_+f_0^*]$ & $\phantom{-}0.03\pm0.42\pm0.12$   \\
\rule[-1.7mm]{0mm}{5mm}$U_{+0}^{+,\R}$
     & Coefficient of $\R[f_+f_0^*]$ & $-0.75\pm0.40\pm0.15$    \\
\rule[-1.7mm]{0mm}{5mm}$U_{+0}^{-,\I}$
     & Coefficient of $\I[f_+f_0^*]\cos(\dmd\dmt)$ & $-0.93\pm0.68\pm0.08$  \\
\rule[-1.7mm]{0mm}{5mm}$U_{+0}^{-,\R}$
     & Coefficient of $\R[f_+f_0^*]\cos(\dmd\dmt)$ & $-0.47\pm0.80\ph{0}\pm0.3\ph{0}$ \\
\rule[-1.7mm]{0mm}{5mm}$I_{+0}^{\I}$
     & Coefficient of $\I[f_+f_0^*]\sin(\dmd\dmt)$ & $-0.1\ph{0}\pm1.1\ph{0}\pm0.3\ph{0}$  \\
\rule[-1.7mm]{0mm}{5mm}$I_{+0}^{\R}$
     & Coefficient of $\R[f_+f_0^*]\sin(\dmd\dmt)$ & $\phantom{-}0.2\ph{0}\pm1.1\ph{0}\pm0.4\ph{0}$ \\[0.15cm]

\rule[-1.7mm]{0mm}{5mm}$U_{-0}^{+,\I}$
     & Coefficient of $\I[f_-f_0^*]$ & $-0.03\pm0.40\pm0.23$ \\
\rule[-1.7mm]{0mm}{5mm}$U_{-0}^{+,\R}$
     & Coefficient of $\R[f_-f_0^*]$ & $-0.52\pm0.32\pm0.08$    \\
\rule[-1.7mm]{0mm}{5mm}$U_{-0}^{-,\I}$
     & Coefficient of $\I[f_-f_0^*]\cos(\dmd\dmt)$ & $\phantom{-}0.24\pm0.61\ph{0}\pm0.2\ph{0}$  \\
\rule[-1.7mm]{0mm}{5mm}$U_{-0}^{-,\R}$
     & Coefficient of $\R[f_-f_0^*]\cos(\dmd\dmt)$ & $-0.42\pm0.73\pm0.28$   \\
\rule[-1.7mm]{0mm}{5mm}$I_{-0}^{\I}$
     & Coefficient of $\I[f_-f_0^*]\sin(\dmd\dmt)$ & $\phantom{-}0.7\ph{0}\pm1.0\ph{0}\pm0.3\ph{0}$  \\
\rule[-1.7mm]{0mm}{5mm}$I_{-0}^{\R}$
     & Coefficient of $\R[f_-f_0^*]\sin(\dmd\dmt)$ & $\phantom{-}0.92\pm0.91\ph{0}\pm0.4\ph{0}$ \\[0.15cm]
\hline
\end{tabular*}
\caption{Fit results for the $U$ and $I$ coefficients. 
	The errors given are statistical (first) and systematic (second).
	The free normalization parameter $U_+^+$ is fixed to 1.}
\label{tab:results}
\end{center}
\end{table*}

\begin{table*}[t]
\centering
\begin{small}
\setlength{\tabcolsep}{0.0pc}
\begin{tabular*}{\textwidth}{@{\extracolsep{\fill}}lccccccccccccc}
\hline
 && \\[-0.3cm]
\rule[-6pt]{0pt}{18pt}  & $\sigPiNb$  &      $\Iz$  &      $\IM$  &   $\IMzIm$  &   $\IMzRe$  &      $\IP$  &   $\IPzIm$  &   $\IPzRe$  &   $\IPMIm$  &   $\IPMRe$  &     $\Uzm$  &     $\Uzp$  &  $\UMzmIm$ \\[0.15cm]
\hline
 &&\\[-0.3cm] $\sigPiNb$  &  $\phantom{-}1.00$  &  &  &  &  &  &  &  &  &  &  &  & \\
      $\Iz$  &  $\phantom{-}0.02$  &  $\phantom{-}1.00$  &  &  &  &  &  &  &  &  &  &  & \\
      $\IM$  &  $\phantom{-}0.04$  &  $-0.07$  &  $\phantom{-}1.00$  &  &  &  &  &  &  &  &  &  & \\
   $\IMzIm$  &  $\phantom{-}0.00$  &  $-0.06$  &  $\phantom{-}0.20$  &  $\phantom{-}1.00$  &  &  &  &  &  &  &  &  & \\
   $\IMzRe$  &  $-0.06$  &  $\phantom{-}0.16$  &  $-0.17$  &  $-0.12$  &  $\phantom{-}1.00$  &  &  &  &  &  &  &  & \\
      $\IP$  &  $-0.02$  &  $-0.01$  &  $-0.08$  &  $-0.13$  &  $\phantom{-}0.12$  &  $\phantom{-}1.00$  &  &  &  &  &  &  & \\
   $\IPzIm$  &  $\phantom{-}0.01$  &  $\phantom{-}0.03$  &  $-0.03$  &  $-0.05$  &  $\phantom{-}0.02$  &  $\phantom{-}0.27$  &  $\phantom{-}1.00$  &  &  &  &  &  & \\
   $\IPzRe$  &  $-0.01$  &  $\phantom{-}0.08$  &  $\phantom{-}0.03$  &  $\phantom{-}0.11$  &  $-0.04$  &  $-0.20$  &  $-0.36$  &  $\phantom{-}1.00$  &  &  &  &  & \\
   $\IPMIm$  &  $\phantom{-}0.07$  &  $\phantom{-}0.01$  &  $\phantom{-}0.10$  &  $-0.05$  &  $\phantom{-}0.01$  &  $\phantom{-}0.29$  &  $\phantom{-}0.10$  &  $-0.07$  &  $\phantom{-}1.00$  &  &  &  & \\
   $\IPMRe$  &  $\phantom{-}0.04$  &  $-0.01$  &  $\phantom{-}0.16$  &  $\phantom{-}0.28$  &  $-0.23$  &  $-0.32$  &  $-0.11$  &  $\phantom{-}0.21$  &  $-0.06$  &  $\phantom{-}1.00$  &  &  & \\
     $\Uzm$  &  $\phantom{-}0.04$  &  $\phantom{-}0.05$  &  $-0.00$  &  $-0.12$  &  $-0.25$  &  $-0.01$  &  $\phantom{-}0.03$  &  $-0.04$  &  $-0.00$  &  $\phantom{-}0.07$  &  $\phantom{-}1.00$  &  & \\
     $\Uzp$  &  $\phantom{-}0.17$  &  $-0.03$  &  $\phantom{-}0.09$  &  $\phantom{-}0.23$  &  $-0.09$  &  $-0.11$  &  $-0.04$  &  $\phantom{-}0.11$  &  $-0.08$  &  $\phantom{-}0.28$  &  $\phantom{-}0.04$  &  $\phantom{-}1.00$  & \\
  $\UMzmIm$  &  $-0.02$  &  $\phantom{-}0.17$  &  $-0.06$  &  $-0.01$  &  $\phantom{-}0.35$  &  $\phantom{-}0.06$  &  $\phantom{-}0.01$  &  $\phantom{-}0.00$  &  $\phantom{-}0.01$  &  $-0.11$  &  $-0.27$  &  $-0.02$  &  $\phantom{-}1.00$ \\
  $\UMzmRe$  &  $\phantom{-}0.01$  &  $-0.05$  &  $-0.05$  &  $-0.41$  &  $-0.09$  &  $\phantom{-}0.01$  &  $\phantom{-}0.01$  &  $\phantom{-}0.03$  &  $\phantom{-}0.02$  &  $\phantom{-}0.02$  &  $\phantom{-}0.11$  &  $-0.08$  &  $\phantom{-}0.02$ \\
  $\UMzpIm$  &  $-0.00$  &  $\phantom{-}0.08$  &  $-0.16$  &  $-0.37$  &  $\phantom{-}0.20$  &  $\phantom{-}0.18$  &  $\phantom{-}0.08$  &  $-0.09$  &  $\phantom{-}0.10$  &  $-0.34$  &  $\phantom{-}0.05$  &  $-0.41$  &  $\phantom{-}0.18$ \\
  $\UMzpRe$  &  $\phantom{-}0.01$  &  $\phantom{-}0.05$  &  $\phantom{-}0.01$  &  $-0.07$  &  $\phantom{-}0.05$  &  $-0.00$  &  $\phantom{-}0.00$  &  $\phantom{-}0.06$  &  $\phantom{-}0.03$  &  $\phantom{-}0.07$  &  $-0.00$  &  $-0.03$  &  $\phantom{-}0.15$ \\
     $\UMm$  &  $\phantom{-}0.06$  &  $\phantom{-}0.01$  &  $-0.01$  &  $-0.03$  &  $-0.02$  &  $-0.01$  &  $\phantom{-}0.01$  &  $\phantom{-}0.02$  &  $\phantom{-}0.01$  &  $\phantom{-}0.07$  &  $-0.01$  &  $-0.03$  &  $\phantom{-}0.09$ \\
     $\UMp$  &  $-0.05$  &  $-0.02$  &  $\phantom{-}0.00$  &  $\phantom{-}0.03$  &  $\phantom{-}0.07$  &  $-0.02$  &  $-0.06$  &  $\phantom{-}0.09$  &  $-0.05$  &  $\phantom{-}0.05$  &  $-0.02$  &  $\phantom{-}0.20$  &  $\phantom{-}0.05$ \\
  $\UPzmIm$  &  $\phantom{-}0.08$  &  $\phantom{-}0.12$  &  $-0.07$  &  $-0.17$  &  $\phantom{-}0.07$  &  $\phantom{-}0.10$  &  $\phantom{-}0.09$  &  $\phantom{-}0.02$  &  $\phantom{-}0.09$  &  $-0.20$  &  $\phantom{-}0.04$  &  $-0.24$  &  $\phantom{-}0.04$ \\
  $\UPzmRe$  &  $\phantom{-}0.06$  &  $\phantom{-}0.06$  &  $\phantom{-}0.03$  &  $\phantom{-}0.07$  &  $-0.13$  &  $-0.09$  &  $-0.15$  &  $\phantom{-}0.09$  &  $\phantom{-}0.00$  &  $\phantom{-}0.20$  &  $\phantom{-}0.09$  &  $\phantom{-}0.04$  &  $-0.07$ \\
  $\UPzpIm$  &  $-0.04$  &  $\phantom{-}0.05$  &  $-0.13$  &  $-0.29$  &  $\phantom{-}0.24$  &  $\phantom{-}0.26$  &  $\phantom{-}0.31$  &  $-0.32$  &  $\phantom{-}0.09$  &  $-0.44$  &  $-0.06$  &  $-0.37$  &  $\phantom{-}0.13$ \\
  $\UPzpRe$  &  $\phantom{-}0.07$  &  $\phantom{-}0.02$  &  $-0.02$  &  $-0.11$  &  $\phantom{-}0.04$  &  $\phantom{-}0.11$  &  $\phantom{-}0.23$  &  $-0.40$  &  $\phantom{-}0.10$  &  $-0.20$  &  $-0.03$  &  $-0.21$  &  $\phantom{-}0.02$ \\
  $\UPMmIm$  &  $-0.05$  &  $-0.05$  &  $-0.01$  &  $\phantom{-}0.07$  &  $-0.02$  &  $-0.07$  &  $-0.04$  &  $-0.03$  &  $-0.06$  &  $-0.09$  &  $-0.04$  &  $\phantom{-}0.03$  &  $-0.02$ \\
  $\UPMmRe$  &  $-0.06$  &  $\phantom{-}0.01$  &  $-0.01$  &  $\phantom{-}0.05$  &  $-0.03$  &  $-0.03$  &  $-0.02$  &  $\phantom{-}0.07$  &  $-0.06$  &  $\phantom{-}0.11$  &  $\phantom{-}0.02$  &  $\phantom{-}0.09$  &  $-0.00$ \\
  $\UPMpIm$  &  $-0.04$  &  $-0.04$  &  $-0.01$  &  $\phantom{-}0.06$  &  $-0.07$  &  $\phantom{-}0.00$  &  $\phantom{-}0.01$  &  $-0.06$  &  $\phantom{-}0.06$  &  $-0.16$  &  $-0.04$  &  $-0.02$  &  $-0.05$ \\
  $\UPMpRe$  &  $-0.10$  &  $-0.07$  &  $\phantom{-}0.03$  &  $\phantom{-}0.06$  &  $\phantom{-}0.01$  &  $-0.04$  &  $-0.02$  &  $-0.07$  &  $\phantom{-}0.03$  &  $-0.11$  &  $-0.07$  &  $-0.00$  &  $\phantom{-}0.01$ \\
     $\UPm$  &  $-0.04$  &  $\phantom{-}0.03$  &  $\phantom{-}0.01$  &  $\phantom{-}0.04$  &  $-0.05$  &  $-0.05$  &  $-0.02$  &  $\phantom{-}0.08$  &  $\phantom{-}0.01$  &  $\phantom{-}0.17$  &  $\phantom{-}0.04$  &  $\phantom{-}0.06$  &  $-0.03$ \\
[0.15cm]\hline
\end{tabular*}

\vspace{1.5\baselineskip}
\setlength{\tabcolsep}{0.0pc}
\begin{tabular*}{\textwidth}{@{\extracolsep{\fill}}lcccccccccccccc}
\hline
 && \\[-0.3cm]
\rule[-6pt]{0pt}{18pt}  &  $\UMzmRe$  &  $\UMzpIm$  &  $\UMzpRe$  &     $\UMm$  &     $\UMp$  &  $\UPzmIm$  &  $\UPzmRe$  &  $\UPzpIm$  &  $\UPzpRe$  &  $\UPMmIm$  &  $\UPMmRe$  &  $\UPMpIm$  &  $\UPMpRe$  &     $\UPm$ \\[0.15cm]
\hline
 &&\\[-0.3cm]  $\UMzmRe$  &  $\phantom{-}1.00$  &  &  &  &  &  &  &  &  &  &  &  &  & \\
  $\UMzpIm$  &  $\phantom{-}0.25$  &  $\phantom{-}1.00$  &  &  &  &  &  &  &  &  &  &  &  & \\
  $\UMzpRe$  &  $\phantom{-}0.25$  &  $\phantom{-}0.12$  &  $\phantom{-}1.00$  &  &  &  &  &  &  &  &  &  &  & \\
     $\UMm$  &  $\phantom{-}0.17$  &  $\phantom{-}0.04$  &  $\phantom{-}0.07$  &  $\phantom{-}1.00$  &  &  &  &  &  &  &  &  &  & \\
     $\UMp$  &  $\phantom{-}0.01$  &  $\phantom{-}0.09$  &  $\phantom{-}0.04$  &  $-0.13$  &  $\phantom{-}1.00$  &  &  &  &  &  &  &  &  & \\
  $\UPzmIm$  &  $\phantom{-}0.04$  &  $\phantom{-}0.24$  &  $\phantom{-}0.00$  &  $\phantom{-}0.01$  &  $-0.07$  &  $\phantom{-}1.00$  &  &  &  &  &  &  &  & \\
  $\UPzmRe$  &  $\phantom{-}0.09$  &  $-0.04$  &  $\phantom{-}0.08$  &  $\phantom{-}0.05$  &  $-0.00$  &  $\phantom{-}0.26$  &  $\phantom{-}1.00$  &  &  &  &  &  &  & \\
  $\UPzpIm$  &  $\phantom{-}0.01$  &  $\phantom{-}0.40$  &  $-0.04$  &  $-0.02$  &  $-0.08$  &  $\phantom{-}0.32$  &  $-0.14$  &  $\phantom{-}1.00$  &  &  &  &  &  & \\
  $\UPzpRe$  &  $-0.03$  &  $\phantom{-}0.09$  &  $-0.02$  &  $\phantom{-}0.02$  &  $-0.26$  &  $\phantom{-}0.17$  &  $-0.09$  &  $\phantom{-}0.35$  &  $\phantom{-}1.00$  &  &  &  &  & \\
  $\UPMmIm$  &  $-0.12$  &  $-0.19$  &  $-0.10$  &  $-0.11$  &  $-0.03$  &  $-0.05$  &  $-0.07$  &  $-0.04$  &  $\phantom{-}0.03$  &  $\phantom{-}1.00$  &  &  &  & \\
  $\UPMmRe$  &  $\phantom{-}0.04$  &  $-0.05$  &  $\phantom{-}0.03$  &  $\phantom{-}0.12$  &  $\phantom{-}0.09$  &  $-0.05$  &  $\phantom{-}0.08$  &  $-0.09$  &  $-0.12$  &  $-0.01$  &  $\phantom{-}1.00$  &  &  & \\
  $\UPMpIm$  &  $-0.09$  &  $-0.17$  &  $-0.13$  &  $-0.03$  &  $-0.10$  &  $-0.02$  &  $-0.05$  &  $-0.04$  &  $\phantom{-}0.08$  &  $\phantom{-}0.35$  &  $-0.02$  &  $\phantom{-}1.00$  &  & \\
  $\UPMpRe$  &  $-0.13$  &  $-0.19$  &  $-0.10$  &  $-0.03$  &  $-0.04$  &  $-0.07$  &  $-0.15$  &  $\phantom{-}0.01$  &  $\phantom{-}0.07$  &  $\phantom{-}0.25$  &  $-0.01$  &  $\phantom{-}0.18$  &  $\phantom{-}1.00$  & \\
     $\UPm$  &  $\phantom{-}0.06$  &  $\phantom{-}0.02$  &  $\phantom{-}0.05$  &  $-0.03$  &  $\phantom{-}0.14$  &  $\phantom{-}0.11$  &  $\phantom{-}0.30$  &  $-0.09$  &  $-0.13$  &  $-0.03$  &  $\phantom{-}0.16$  &  $-0.10$  &  $-0.11$  &  $\phantom{-}1.00$ \\
[0.15cm]\hline
\end{tabular*}

\end{small}
\caption{\label{tab:corrmat}
  Correlation matrix of statistical uncertainties for the $U$
  and $I$ coefficients. Note that
  all elements above the diagonal are omitted for readability.
}
\end{table*}

\begin{figure*}[p]
  \centerline{  \epsfxsize8.2cm\epsffile{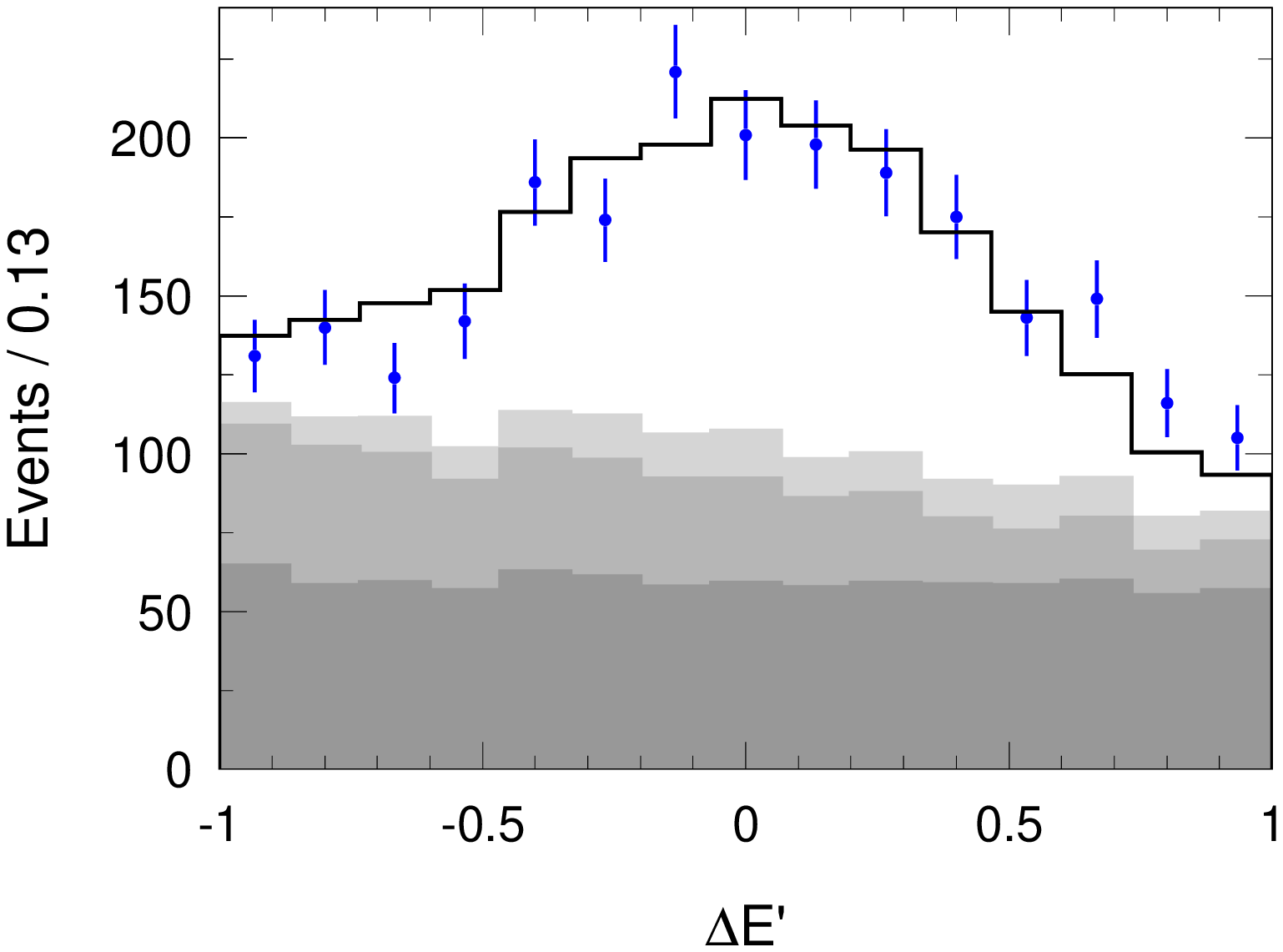}
                \epsfxsize8.2cm\epsffile{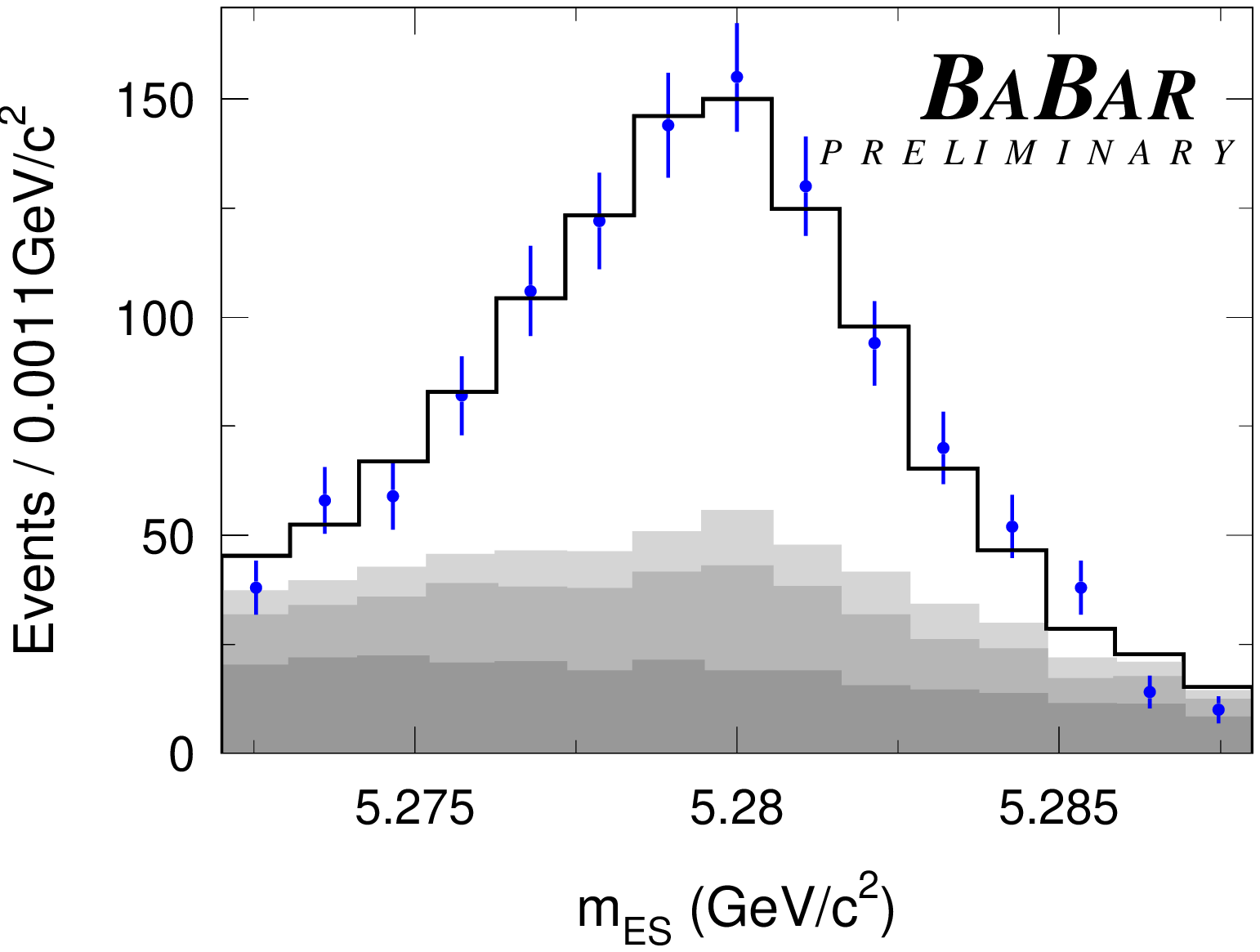}}
  \vspace{-0.1cm}
  \centerline{  \epsfxsize8.2cm\epsffile{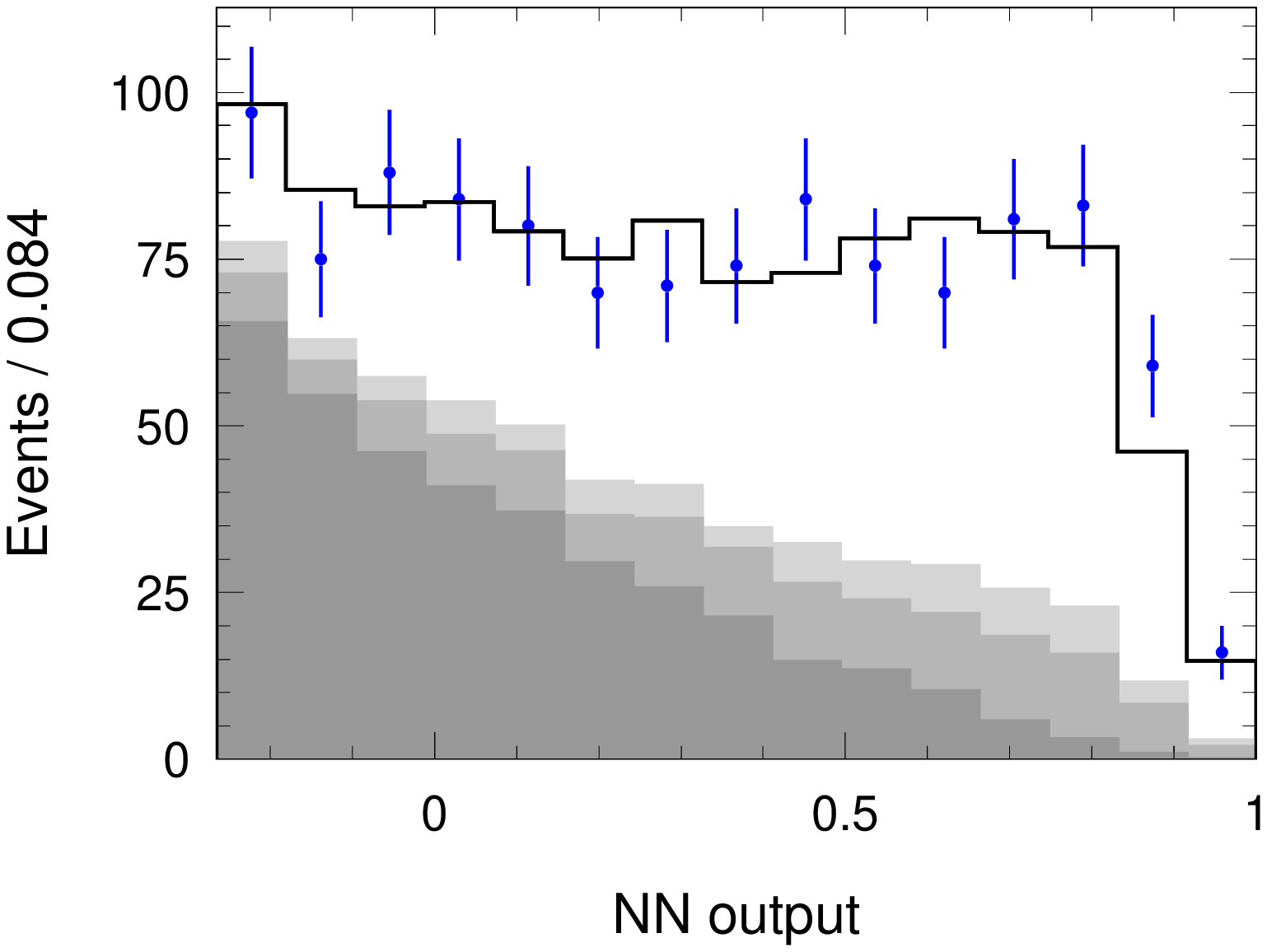} 
                \epsfxsize8.2cm\epsffile{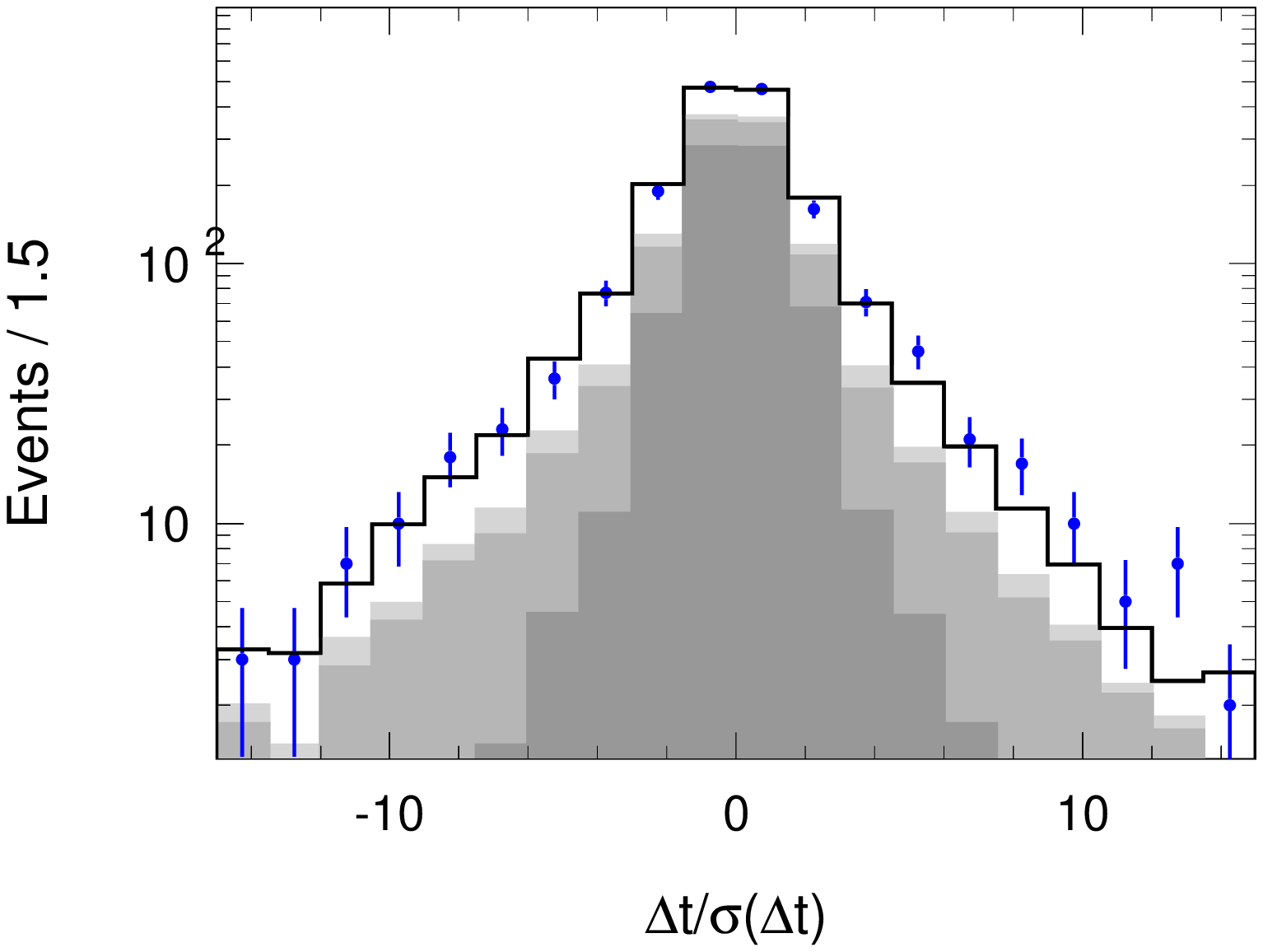}}
  \vspace{-0.1cm}
  \centerline{  \epsfxsize8.2cm\epsffile{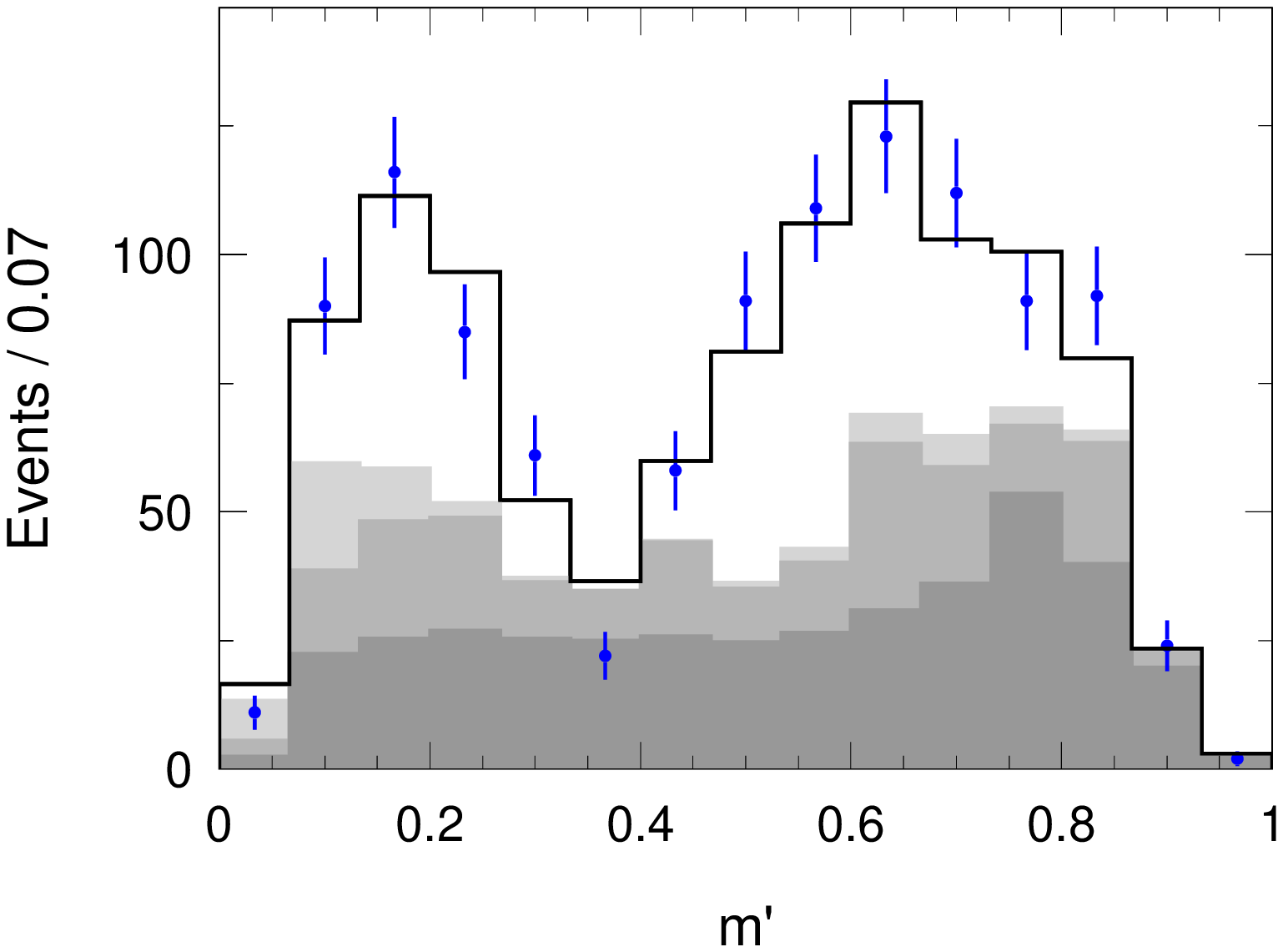}
                \epsfxsize8.2cm\epsffile{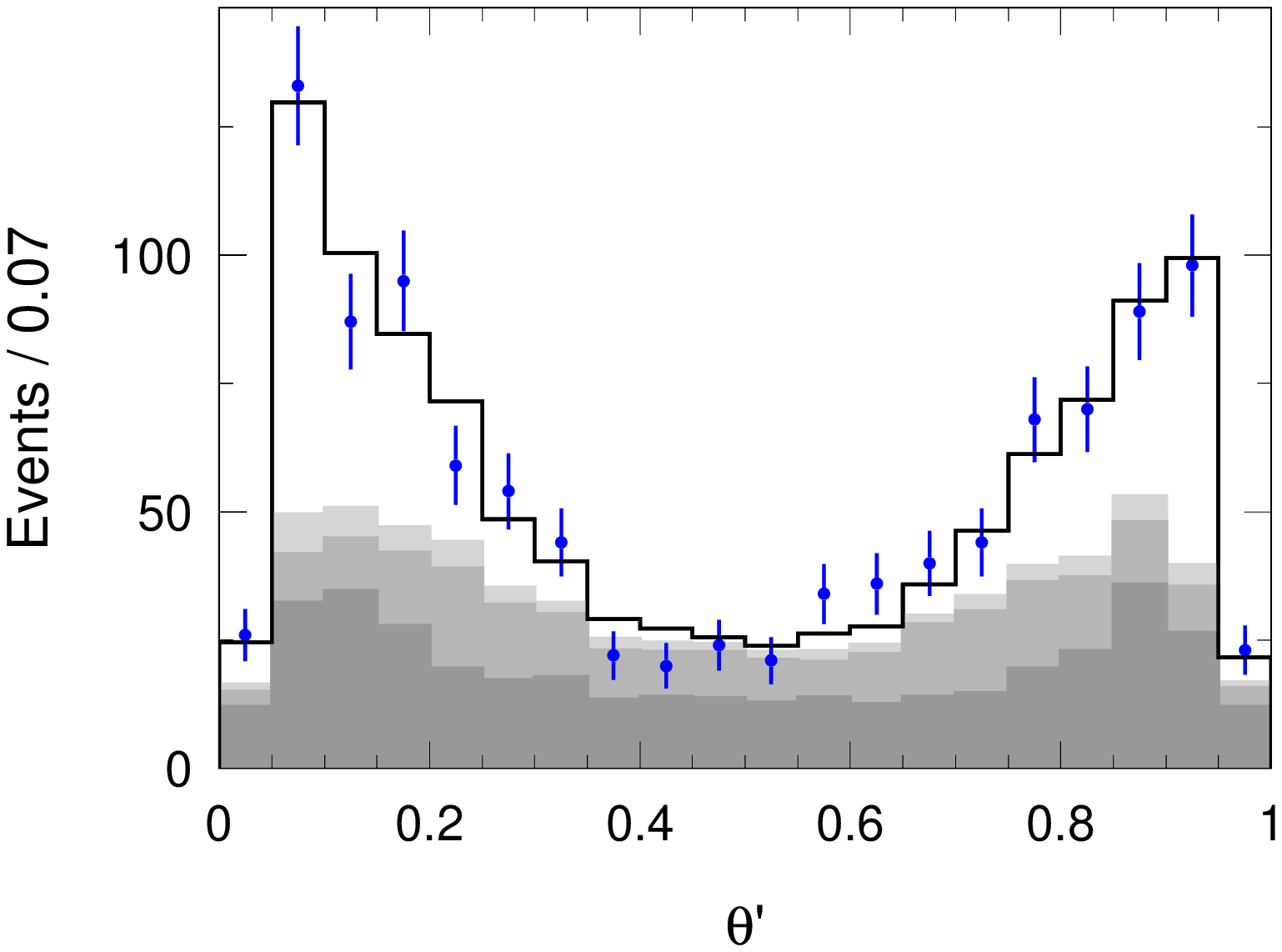}}
  \vspace{-0.5cm}
  \caption{\label{fig:projections} 
	Distributions of (top to bottom, left to right) $\deprime$, $\mes$, 
	NN output, $\dt/\sigma(\dt)$,  $\mprime$ and $\thetaprime$ for samples 
	enhanced in $\Btopipipi$ signal. The dots with error bars give 
	the on-resonance data. The solid histogram shows the
	projection of the fit result. The dark,
	medium and light shaded areas represent respectively the contribution
	from continuum events, the sum of continuum events 
	and the $B$-background expectation, and the sum of these and 
	the misreconstructed signal events. }
\end{figure*}

\begin{figure}[t]
  \centerline{\epsfxsize12.cm\epsffile{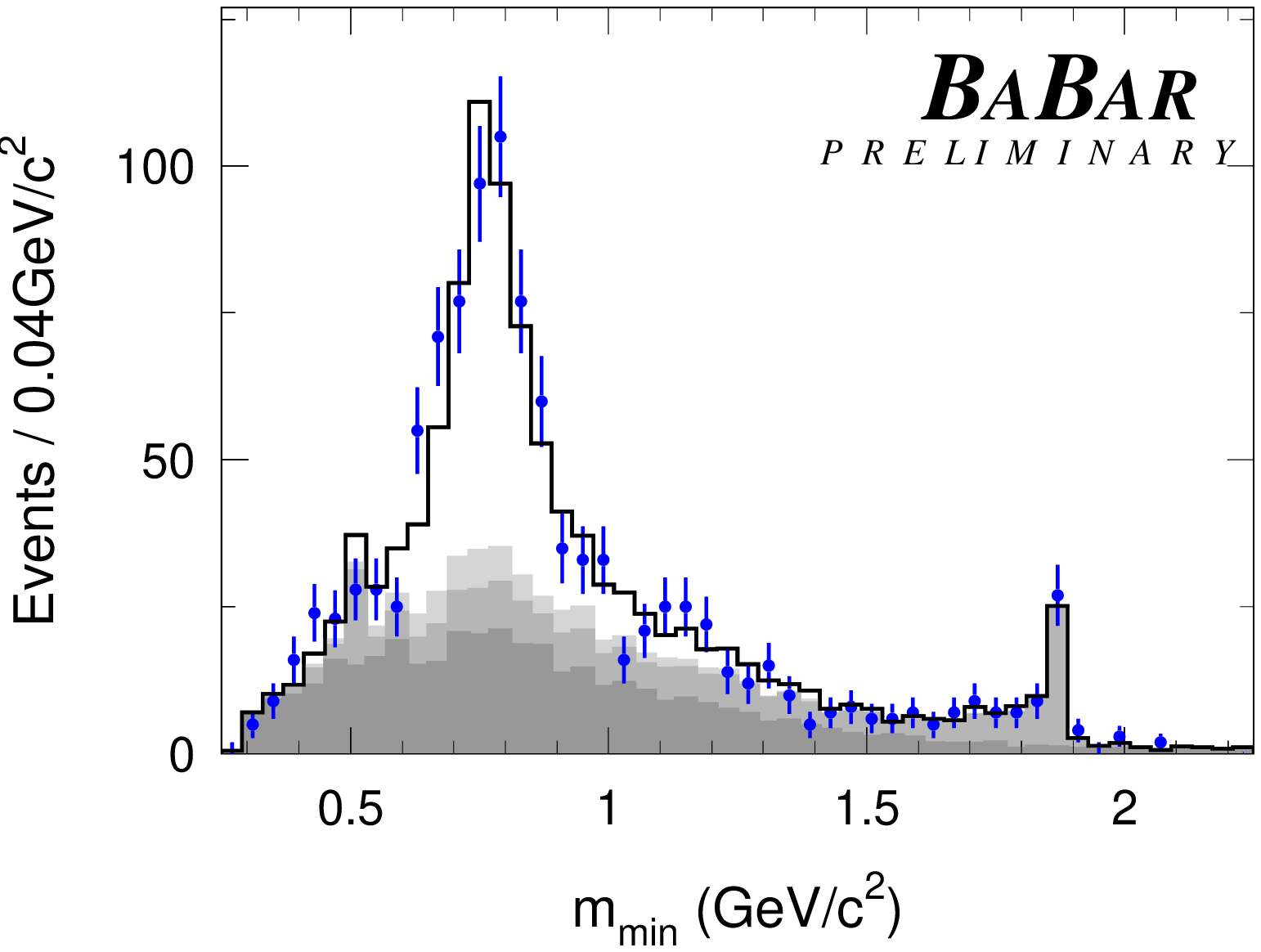}}
  \vspace{-0.5cm}
  \caption{\label{fig:projection_mmin} 
	Distribution of minimum of the three di-pion invariant masses,
        for samples enhanced in $\Btopipipi$ signal. The dots with error bars give 
	the on-resonance data. The solid histogram shows the
	projection of the fit result. The dark,
	medium and light shaded areas represent respectively the contribution
	from continuum events, the sum of continuum events 
	and the $B$-background expectation, and the sum of these and 
	the misreconstructed signal events. }
\end{figure}

The maximum-likelihood fit results in the $\Btopipipi$ event yield
$1847\pm69$, where the error is statistical only. For the $U$ and $I$
coefficients, the results are given together with their statistical
and systematic errors in Table~\ref{tab:results}. The corresponding 
correlation matrix is given in Table~\ref{tab:corrmat}. We have 
generated a sample of Monte Carlo experiments to determine the probability
density distributions of the fit parameters. 
Within the statistical 
uncertainties of this sample we find Gaussian distributions of the 
distribution for the fitted $U$ and $I$ coefficients. 
This allows us to 
use the least-squares method to derive other quantities from these
(Section~\ref{sec:Physics}).

The signal is 
dominated by $\Bz\to\rho^\pm\pi^\mp$ decays. We observe an excess of
$\rho^0\piz$ events, which is in agreement with our previous upper
limit~\cite{BABARrho0pi0}, and the latest measurement from the Belle 
collaboration~\cite{BELLErho0pi0}.
The result for the $\rho(1450)$ amplitude is in agreement with the findings 
in $\tau$ and $\epem$ decays~\cite{taueeref}. For the relative strong phase between 
the $\rho(770)$ and the $\rho(1450)$ amplitudes we find 
$(171\pm23)^\circ$ (statistical error only), which is 
compatible with the result from $\tau$ and $\epem$ data.

Figure~\ref{fig:projections} shows distributions 
of $\deprime$, $\mes$, the NN output, $\dt/\sigma(\dt)$, where $\sigma(\dt)$
is the per-event error on $\dt$, as well as the Dalitz plot 
variables $\mprime$ and $\thetaprime$,
which are enhanced in signal content by requirements on the signal-to-continuum 
likelihood ratios of the other discriminating variables. 
Figure~\ref{fig:projection_mmin} shows distribution
of the minimum of three di-pion invariant masses, again
enhanced in signal content. This plot shows clearly that
$\rho(770)$ dominates the signal component.

As a validation of our treatment of the time dependence  we allow
$\tau_{\Bz}$ to vary in the fit. We find
$\tau_{\Bz} = (1.513\pm 0.066)\ps$,
while the remaining free parameters are consistent with the nominal fit.
To validate the SCF modeling, we leave the average SCF fractions per tagging
category free to vary in the fit and find results that are consistent
with the MC prediction. 

\section{INTERPRETATION OF THE RESULTS}
\label{sec:Physics}

The $U$ and $I$ coefficients are related to the 
quasi-two-body parameters, defined in Ref.~\cite{rhopipaper}
as follows
\beqn
\label{eq:q2bparams}
	C^+ = \frac{ U^-_+ }{ U^+_+ }~, \hspace{0.6cm}
	C^- = \frac{ U^-_- }{ U^+_- }~, \hspace{0.6cm}
	S^+ = \frac{ 2 \, I_+ }{ U^+_+ }~, \hspace{0.6cm}
	S^- = \frac{ 2 \, I_- }{ U^+_- }~, \hspace{0.6cm}
	\Acp = \frac{ U^+_+ \, - U^+_- }{ U^+_+ \, + U^+_- }~, 
\eeqn
and where $C=(C^++C^-)/2$, $\dC=(C^+-C^-)/2$, $S=(S^++S^-)/2$, 
and $\dS=(S^+-S^-)/2$ and $\Acp$ is the time and flavor integrated asymmetry.
 In contrast to our previous 
analysis~\cite{rhopipaper}, the definitions of Eq.~(\ref{eq:q2bparams})
explicitly account for the presence of interference effects, and are 
thus exact even for a $\rho$ with finite width, as long as the $U$ and 
$I$ coefficients are obtained with a Dalitz plot analysis. This treatment
leads to a dilution of the result and hence to slightly
increased statistical uncertainties compared to neglecting
the interference effects.

For the \CP-violation parameters, we obtain
\beqn
	\Acp    &=&      -0.142\pm 0.041 \pm{0.015}~, \\
	C   	&=& \ph{-}0.154 \pm 0.090 \pm 0.037~, \\
	S 	&=& \ph{-}0.01\pm 0.12\pm 0.028~,
\eeqn
where the first errors given are statistical and the second 
are the systematic uncertainties.
For the other parameters in the quasi-two-body description of the 
$\Bz(\Bzb) \to \rhopi$ decay-time dependence, we measure
\beqn
	\dC 	&=& 0.377\pm 0.091\pm 0.021~, \\
	\dS 	&=& 0.06\pm 0.13\pm 0.029~.
\eeqn
The systematic errors are dominated by the 
uncertainty on the \CP content of the \B-related backgrounds.
Other contributions are the signal description in the likelihood 
model (including the limit on non-resonant $\Btopipipi$ events), and
the fit bias uncertainty. The covariance matrix, including
systematic, of the five quasi-two-body parameters is given
in Table~\ref{tab:q2bCov}.

\begin{table*}[t]
\begin{center}
\setlength{\tabcolsep}{0.0pc}
\begin{tabular*}{\textwidth}{@{\extracolsep{\fill}}lccccc}
\hline
 && \\[-0.3cm]
\rule[-6pt]{0pt}{18pt}  &     $\Acp$  &       $C$  &      $\dC$  &       $S$  &      $\dS$ \\[0.15cm]
\hline
 &&\\[-0.3cm]     
     $\Acp$  &  $\phantom{-}1.93$  &  &  &  & \\
        $C$  &  $-0.71$  &  $\phantom{-}9.68$  &  &  & \\
      $\dC$  &  $-0.55$  &  $\phantom{-}2.63$  &  $\phantom{-}8.93$  &  & \\
        $S$  &  $-0.03$  &  $-0.71$  &  $-0.13$  &  $\phantom{-}15.3$  & \\
      $\dS$  &  $-0.03$  &  $-0.57$  &  $-0.07$  &  $\phantom{-}3.93$  &  $\phantom{-}17.08$ \\
[0.15cm]\hline
\end{tabular*}
\caption{ \label{tab:q2bCov}
        Covariance matirx multipled by 1000 of the quasi-two-body parameters.}
\end{center}
\end{table*}

One can transform the experimentally convenient, namely
uncorrelated, direct \CP-violation parameters $\Crhopi$ and $\Acp$
into the physically more intuitive quantities $\Acppm$, $\Acpmp$,
defined by
\beqn
\label{eq:Adirpm}
    \Acppm &=& \frac{|\kappm|^2-1}{|\kappm|^2+1}
    \;=\; -\frac{\Acp+\Crhopi+\Acp\dCrhopi}{1+\dCrhopi+\Acp\Crhopi}
    ~,\\[0.2cm]\nonumber
\label{eq:Adirmp}
    \Acpmp &=& \frac{|\kapmp|^2-1}{|\kapmp|^2+1}
    \;=\; \frac{\Acp-\Crhopi-\Acp\dCrhopi}{1-\dCrhopi-\Acp\Crhopi}
    ~,
\eeqn
where
$\kappm = (q/p)(\Ampb/\Apm)$ and $\kapmp = (q/p)(\Apmb/\Amp)$,
so that $\Acppm$ ($\Acpmp$) involves only diagrams where the $\rho$
($\pi$) meson is emitted by the $W$ boson. We find
\begin{figure}[t]
  \centerline{  \epsfxsize7.7cm\epsffile{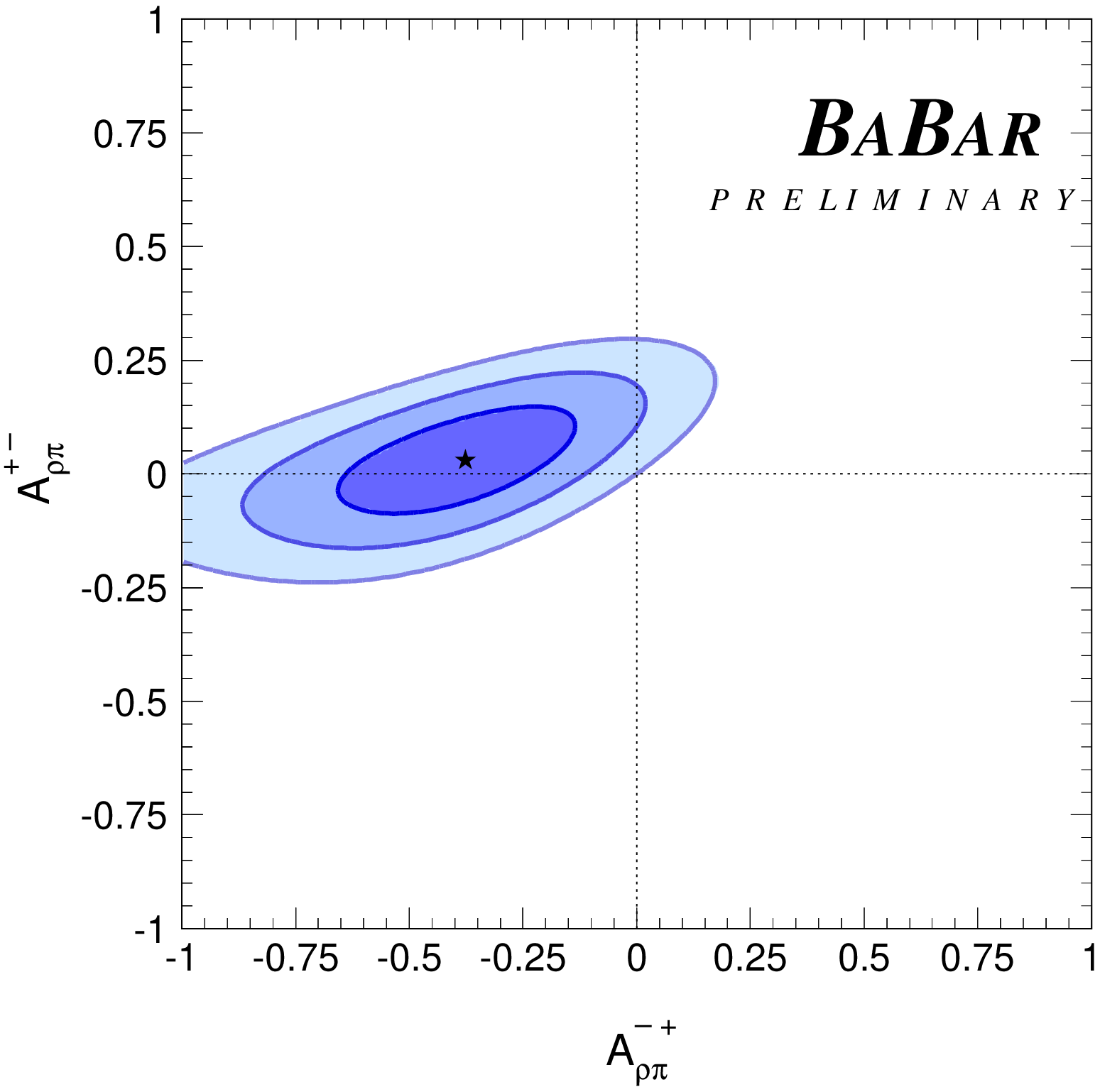}}
  \vspace{-0.1cm}
  \caption{\label{fig:apmamp} 
	Confidence level contours for the direct \CP asymmetries
	$\Acppm$ versus $\Acpmp$. The shaded areas represent 
	$1\sigma$, $2\sigma$ and $3\sigma$ contours, respectively. }
\end{figure}
\beqn
    \Acppm &=& 0.03\pm0.07\pm0.03~, \\
    \Acpmp &=& -0.38^{\,+0.15}_{\,-0.16}\pm0.07~,
\eeqn
with a correlation coefficient of 0.62 between $\Acppm$
and $\Acpmp$. The confidence level contours including systematic
errors are given in Fig.~\ref{fig:apmamp}. The significance, including 
systematic uncertainties and calculated by using a mininum $\chi^2$
method, for the observation of non-zero direct \CP violation is 
at the $3.0\sigma$ level. The evidence of direct \CP violation is
almost entirely from the $\Bz\to\rho^-\pi^+$ decays.
\begin{figure*}[t]
  \centerline{  \epsfxsize8.2cm\epsffile{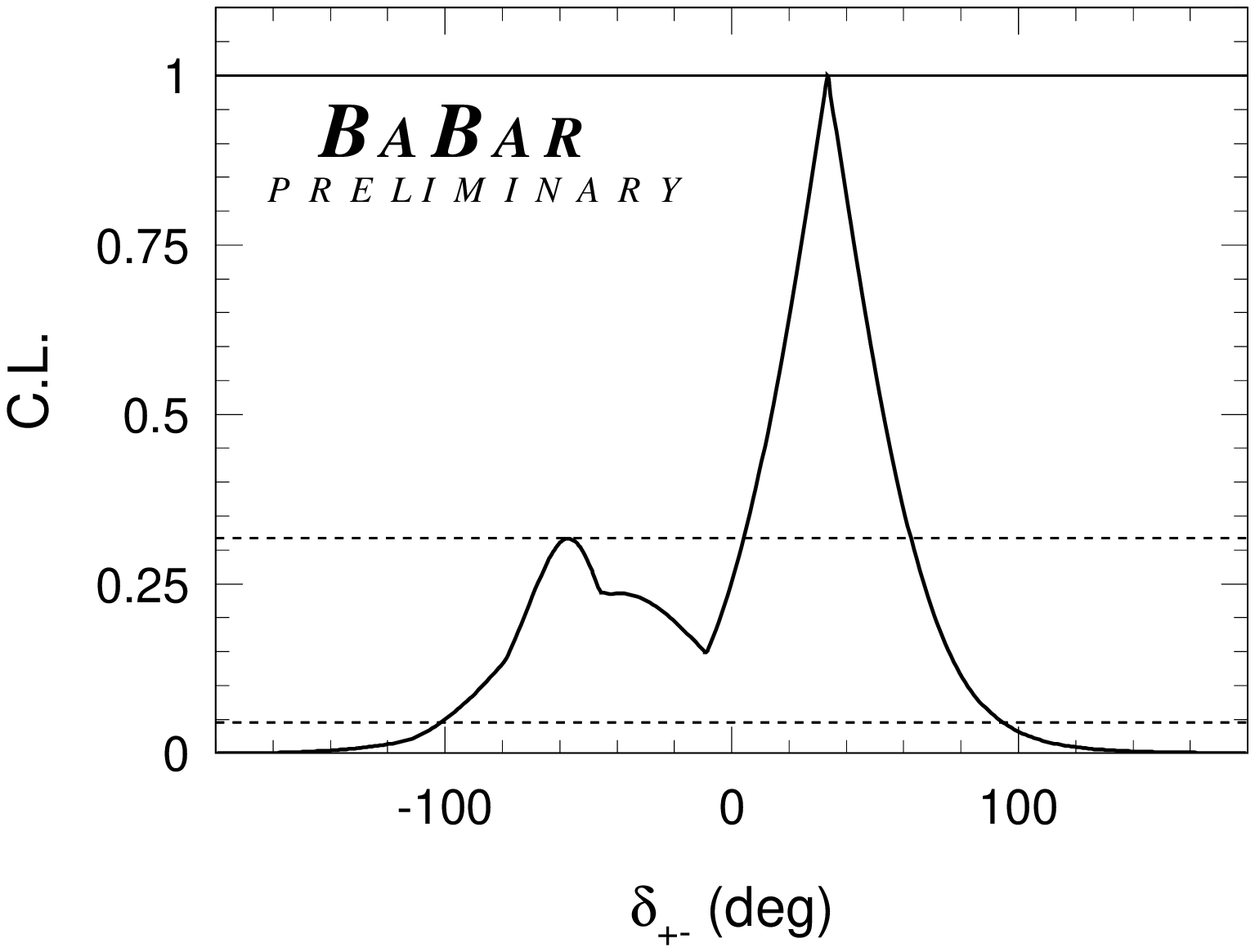}
   	        \epsfxsize8.2cm\epsffile{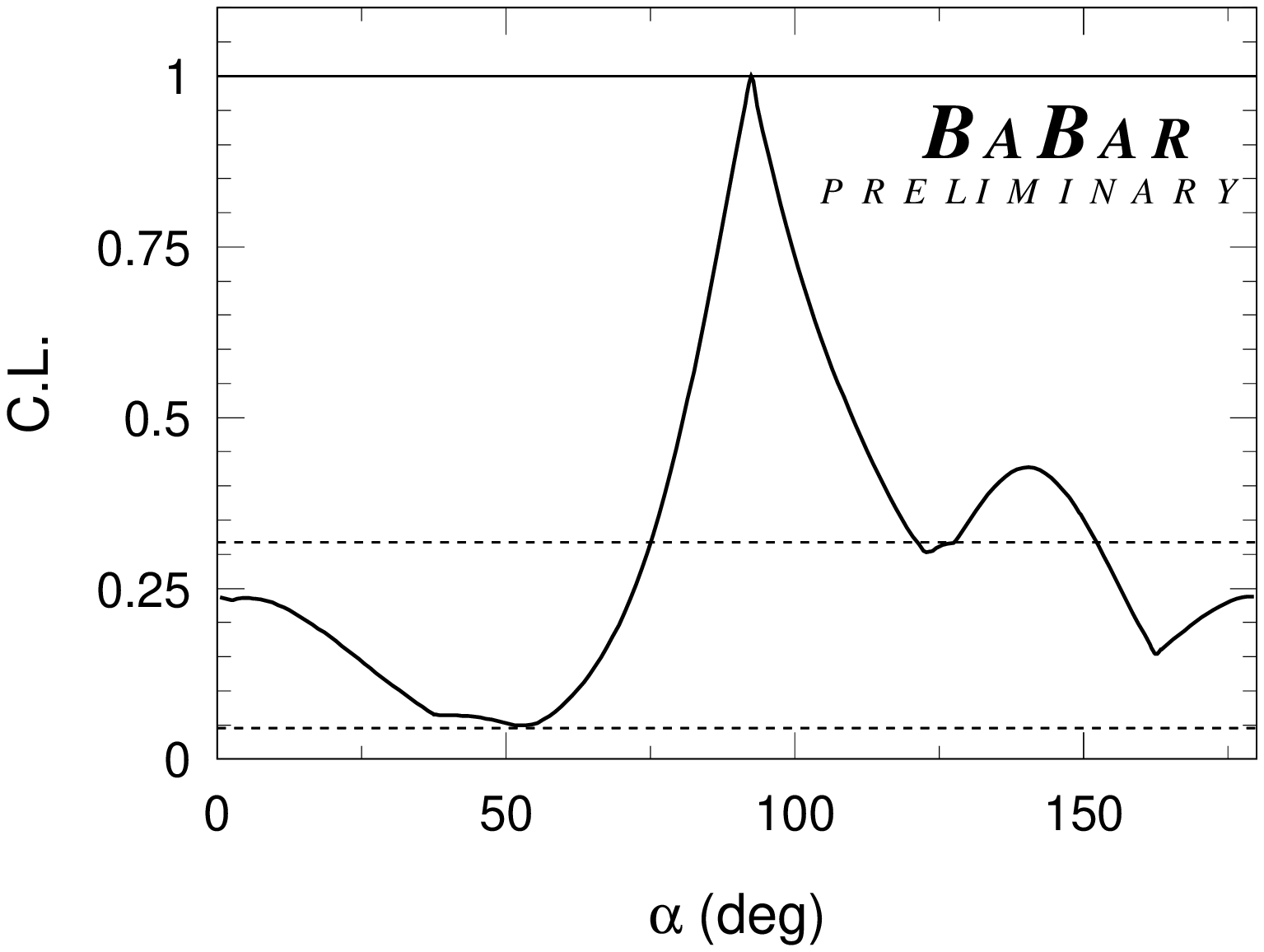}}
  \vspace{-0.4cm}
  \caption{\label{fig:deltaalpha} 
	Confidence level functions for $\delta_{+-}$ (left) and 
	$\alpha$ (right). Indicated 
	by the dashed horizontal lines are the confidence level (C.L.) values
        corresponding to $1\sigma$ and $2\sigma$, respectively.}
\end{figure*}

The measurement of the resonance interference terms allows us to 
determine the relative phase
\beq
\label{eq:deltapm}
\delta_{+-} = \arg\left( A^{+*}A^{-} \right)~,
\eeq
between the amplitudes of the 
decays $B^0\to\rho^-\pi^+$ and $B^0\to\rho^+\pi^-$. Through 
the definitions~(\ref{eq:firstObs})--(\ref{eq:lastObs}), we can derive 
a constraint on $\delta_{+-}$ from the measured $U$ and $I$ 
coefficients
by performing a least-squares minimization with the six complex amplitudes
as free parameters. 
The constraint can be improved with the use of strong 
isospin symmetry. The amplitudes 
$\Aij$ represent the sum of tree-level and penguin-type
amplitudes, which have different CKM factors: the tree-level $(\Tij)$
$\Bz\to\rho^\kappa\pi^{\kappab}$ transition amplitude
is proportional to 
$V_{ud}V_{ub}^*$, while the corresponding penguin-type amplitude $(\Pij)$
comes with $V_{qd}V_{qb}^*$, where $q=u,c,t$. 
Here we denote by $\kappab$ the charge conjugate of $\kappa$,
where $\overline 0=0$. 
Using the unitarity of the 
CKM matrix one can reorganize the amplitudes and obtain~\cite{BaBarPhysBook}
\beqn
\label{eq:aijamps}
	\Aij 		&=& \Tij e^{-i\alpha} + \Pij ~, \nonumber\\
	\Abij 		&=& \Tji e^{+i\alpha} + \Pji ~,
\eeqn
where the magnitudes of the CKM factors have been absorbed in the 
$\Tij$, $\Pij$, $\Tji$ and $\Pji$. 
The Eqs.~(\ref{eq:aijamps}) represent 13 unknowns
of which two can be fixed due to an arbitrary
global phase and the normalization condition $U_+^+=1$. Using
strong isospin symmetry and neglecting isospin-breaking effects,
one can identify $P^{0}=-(P^{+}+P^{-})/2$, which reduces the 
number of unknowns to be determined by the fit to 9. This set of parameters 
provides the constraint on $\delta_{+-}$, shown in the 
left plot of Fig.~\ref{fig:deltaalpha}. 
We find for the solution that is favored by the fit
\beq
	\delta_{+-} \; = \; \left(34\,\pm29\right)^\circ~,
\eeq
where the errors include both statistical and systematic effects.
There is a clear structure of multiple solutions which give comparable
$\chi^2$. 
Only a marginal constraint on $\delta_{+-}$ is obtained for ${\rm C.L.}<0.05$. 

Finally, following the same procedure, we can also
derive a constraint on $\alpha$ from the measured $U$ and $I$ 
coefficients. The resulting
C.L. function versus $\alpha$ is given in the right hand plot
of Fig.~\ref{fig:deltaalpha}. It includes systematic uncertainties.
Ignoring the mirror solution at $\alpha + 180^\circ$, we find
$\alpha \; \in \; (75^\circ, 152^\circ)$ at $68\%$ C.L. 
No constraint on $\alpha$ is achieved at two sigma and beyond.

\section{SUMMARY}
\label{sec:Summary}

We have presented the preliminary measurement of 
\CP-violating asymmetries in $\Btopipipi$ decays dominated by 
the $\rho$ resonance. The results are obtained from a data sample 
of 347 million $\FourS \to B\Bbar$ decays. We perform a time-dependent 
Dalitz plot analysis. From the measurement of the coefficients of 26 form 
factor bilinears we determine the three \CP-violating 
and two \CP-conserving quasi-two-body parameters, where we find a 
$3.0\sigma$ evidence of direct \CP violation. Taking advantage of 
the interference between the $\rho$ resonances in the Dalitz plot,
we derive constraints on the relative strong phase between 
$\Bz$ decays to $\rho^+\pim$ and $\rho^-\pip$, and on the angle
$\alpha$ of the Unitarity Triangle. These measurements are 
consistent with the expectation from the CKM fit~\cite{alphaSM}.

\section{Acknowledgments}
\label{sec:acknowledgments}
We are grateful for the 
extraordinary contributions of our \pep2\ colleagues in
achieving the excellent luminosity and machine conditions
that have made this work possible.
The success of this project also relies critically on the 
expertise and dedication of the computing organizations that 
support \babar.
The collaborating institutions wish to thank 
SLAC for its support and the kind hospitality extended to them. 
This work is supported by the
US Department of Energy
and National Science Foundation, the
Natural Sciences and Engineering Research Council (Canada),
Institute of High Energy Physics (China), the
Commissariat \`a l'Energie Atomique and
Institut National de Physique Nucl\'eaire et de Physique des Particules
(France), the
Bundesministerium f\"ur Bildung und Forschung and
Deutsche Forschungsgemeinschaft
(Germany), the
Istituto Nazionale di Fisica Nucleare (Italy),
the Foundation for Fundamental Research on Matter (The Netherlands),
the Research Council of Norway, the
Ministry of Science and Technology of the Russian Federation, 
Ministerio de Educaci\'on y Ciencia (Spain), and the
Particle Physics and Astronomy Research Council (United Kingdom). 
Individuals have received support from 
the Marie-Curie IEF program (European Union) and
the A. P. Sloan Foundation.

\end{document}